\documentclass[acmlarge]{acmart}

\usepackage[utf8]{inputenc}
\usepackage[T1]{fontenc}
\usepackage{graphicx}
\usepackage{adjustbox}
\usepackage{tabularx}
\usepackage{subcaption}
\usepackage[english]{babel}
\usepackage[figuresright]{rotating}
\usepackage{makecell}
\usepackage{array}
\usepackage{caption}
\usepackage{multirow}
\usepackage{url}
\usepackage{float}
\usepackage{hyperref}
\usepackage{adjustbox}
\usepackage{threeparttable}
\usepackage{rotating} 
\usepackage{array}
\usepackage{makecell}
\usepackage{tikz}
\usepackage{afterpage}
\usepackage{longtable,lscape}
\usepackage{enumitem}
\usepackage{color, colortbl}
\usepackage{makecell}
\usepackage{wrapfig}
\usepackage[linesnumbered]{algorithm2e}
\definecolor{Gray}{gray}{0.9}
\definecolor{LightGray}{gray}{0.6}
\usepackage[most]{tcolorbox}
\hypersetup{colorlinks=true, citecolor={magenta}, filecolor=blue, linkcolor=black, urlcolor=blue}

\usepackage{xcolor}
\definecolor{green(munsell)}{rgb}{0.0, 0.66, 0.47}
\definecolor{cadmiumgreen}{rgb}{0.0, 0.42, 0.24}
\definecolor{cobalt}{rgb}{0.0, 0.28, 0.67}
\definecolor{amber(sae/ece)}{rgb}{1.0, 0.49, 0.0}
\newlength\MAX  

\newlength\BARSIZE  

\newcommand*\ChartBarBlue[1]{\textcolor{cobalt}{\rule{\BARSIZE}{2ex}}}
\newcommand*\ChartBarGreen[1]{\textcolor{green(munsell)}{\rule{\BARSIZE}{2ex}}}
\newcommand*\ChartBarOrange[1]{\textcolor{amber(sae/ece)}{\rule{\BARSIZE}{2ex}}}

\definecolor{red}{rgb}{0, 0, 0}
\definecolor{Gray}{gray}{0.9}
\definecolor{Gray2}{gray}{0.8}

\newcolumntype{a}{>{\columncolor{Gray}}c}
\newcolumntype{b}{>{\columncolor{white}}c}

\AtBeginDocument{%
  \providecommand\BibTeX{{%
    \normalfont B\kern-0.5em{\scshape i\kern-0.25em b}\kern-0.8em\TeX}}}

\setcopyright{acmlicensed}
\acmJournal{IMWUT}
\acmYear{2022} \acmVolume{6} \acmNumber{4} \acmArticle{176} \acmMonth{12} \acmPrice{15.00}\acmDOI{10.1145/3569483}

\begin{document}

\title{Generalization and Personalization of Mobile Sensing-Based Mood Inference Models: An Analysis of College Students in Eight Countries}

\author{Lakmal Meegahapola}
\email{lmeegahapola@idiap.ch}
\orcid{0000-0002-5275-6585}
\affiliation{
\institution{Idiap Research Institute \& EPFL} \country{Switzerland}}

\author{William Droz}
\orcid{0000-0003-0379-2018}
\affiliation{\institution{Idiap Research Institute} \country{Switzerland}}

\author{Peter Kun}
\orcid{0000-0003-0778-7662}
\author{Amalia de Götzen}
\orcid{0000-0001-7214-5856}
\affiliation{\institution{Aalborg University} \country{Denmark}}

\author{Chaitanya Nutakki}
\orcid{0000-0002-5164-2391}
\author{Shyam Diwakar}
\orcid{0000-0003-1546-0184}
\affiliation{\institution{Amrita Vishwa Vidyapeetham} \country{India}}

\author{Salvador Ruiz Correa}
\orcid{0000-0002-2918-6780}
\affiliation{\institution{Instituto Potosino de Investigación Científica y Tecnológica} \country{Mexico}}

\author{Donglei Song}
\orcid{0000-0001-6737-6932}
\author{Hao Xu}
\orcid{0000-0001-8474-0767}
\affiliation{\institution{Jilin University} \country{China}}

\author{Miriam Bidoglia}
\orcid{0000-0002-1583-6551}
\author{George Gaskell}
\orcid{0000-0001-6135-9496}
\affiliation{\institution{London School of Economics and Political Science} \country{UK}}

\author{Altangerel Chagnaa}
\orcid{0000-0003-2331-3045}
\author{Amarsanaa Ganbold}
\orcid{0000-0003-4335-6608}
\author{Tsolmon Zundui}
\orcid{0000-0002-2797-517X}
\affiliation{\institution{National University of Mongolia} \country{Mongolia}}

\author{Carlo Caprini}
\orcid{0000-0001-6609-1572}
\author{Daniele Miorandi}
\orcid{0000-0002-3089-977X}
\affiliation{\institution{U-Hopper} \country{Italy}}

\author{Alethia Hume}
\orcid{0000-0002-1874-1419}
\author{Jose Luis Zarza}
\orcid{0000-0003-0069-4287}
\author{Luca Cernuzzi}
\orcid{0000-0001-7803-1067}
\affiliation{\institution{Universidad Católica "Nuestra Señora de la Asunción"} \country{Paraguay}}

\author{Ivano Bison}
\orcid{0000-0002-9645-8627}
\author{Marcelo Rodas Britez}
\orcid{0000-0002-7607-7587}
\author{Matteo Busso}
\orcid{0000-0002-3788-0203}
\author{Ronald Chenu-Abente}
\orcid{0000-0002-1121-0287}
\author{Can Günel}
\orcid{0000-0002-8750-6498}
\author{Fausto Giunchiglia}
\orcid{0000-0002-5903-6150}
\affiliation{\institution{University of Trento} \country{Italy}}

\author{Laura Schelenz}
\orcid{0000-0003-3320-2314}
\affiliation{\institution{University of Tübingen} \country{Germany}}

\author{Daniel Gatica-Perez}
\orcid{0000-0001-5488-2182}
\email{gatica@idiap.ch}
\affiliation{\institution{Idiap Research Institute \& EPFL} \country{Switzerland}}

\renewcommand{\shortauthors}{Meegahapola et al.}
\renewcommand{\shorttitle}{Generalization and Personalization of Mobile Sensing-Based Mood Inference Models}

\begin{abstract}

Mood inference with mobile sensing data has been studied in ubicomp literature over the last decade. This inference enables context-aware and personalized user experiences in general mobile apps and valuable feedback and interventions in mobile health apps. However, even though model generalization issues have been highlighted in many studies, the focus has always been on improving the accuracies of models using different sensing modalities and machine learning techniques, with datasets collected in homogeneous populations. In contrast, less attention has been given to studying the performance of mood inference models to assess whether models generalize to new countries. In this study, we collected a mobile sensing dataset with 329K self-reports from 678 participants in eight countries (China, Denmark, India, Italy, Mexico, Mongolia, Paraguay, UK) to assess the effect of geographical diversity on mood inference models. We define and evaluate country-specific (trained and tested within a country), continent-specific (trained and tested within a continent), country-agnostic (tested on a country not seen on training data), and multi-country (trained and tested with multiple countries) approaches trained on sensor data for two mood inference tasks with population-level (non-personalized) and hybrid (partially personalized) models. We show that partially personalized country-specific models perform the best yielding area under the receiver operating characteristic curve (AUROC) scores of the range 0.78-0.98 for two-class (negative vs. positive valence) and 0.76-0.94 for three-class (negative vs. neutral vs. positive valence) inference. Further, with the country-agnostic approach, we show that models do not perform well compared to country-specific settings, even when models are partially personalized. We also show that continent-specific models outperform multi-country models in the case of Europe. Overall, we uncover generalization issues of mood inference models to new countries and how the geographical similarity of countries might impact mood inference.

\end{abstract}



\maketitle

\section{Introduction}\label{sec:introduction}

Mental well-being related issues are common among young adults due to a plethora of personal and societal reasons such as leaving home, study workload, poor financial stability, and complex social relationships \cite{patel2007mental, rickwood2007and}. These issues are even more prominent in the post-pandemic world, where social relationships have taken a toll due to more emphasis on remote work/study settings. Some studies have shown that this emerging lifestyle has affected phone usage behavior as well \cite{stockwell_changes_2021, zheng_covid-19_2020, ratan_smartphone_2021, saadeh_smartphone_2021, li_impact_2021}. Further, declining mental well-being conditions could lead to adverse outcomes such as substance abuse and suicidal thoughts \cite{rowe2003substance, crosby2011suicidal, franklin2017risk}. In this context, prior research has discussed the potential of timely and accurate mood tracking for both personal and clinical care \cite{spathis2019passive, wang2016crosscheck, faherty2017pregnancy, mark2008chart}. Ecological momentary assessments (EMAs) and survey questionnaires are commonly used for mood tracking. However, such techniques are burdensome to users, and prior work has shown that it is difficult to sustain the practice of reporting for long periods unless there is a strong motivation \cite{baumel2019objective, rapp2014self, schueller2021understanding}. As a possible alternative, multi-modal sensors in smartphones could be used to infer mood unobtrusively with reasonable accuracies \cite{pratap2019accuracy, likamwa2013moodscope, servia2017mobile}. 

According to prior work in psychology and social sciences, physiological aspects, including mood, are perceived and expressed differently in different countries, cultures, and societies \cite{luomala2004cross} \footnote{For pragmatic reasons, we are equating the geographical location (country) of our participants with a specific culture that is distinct to this particular country. We acknowledge that cultures can be multidimensional and exist in tension with each other and in plurality within the same geographic boundary \cite{yuval2004gender}. However, throughout the paper, we use country, culture, and geographic region interchangeably.}. According to a cross-country study by Becht et al. \cite{becht2002crying}, mood and related behaviors could vary based on a person's culture, and perceptions and beliefs regarding different moods stemming from one's culture. However, prior work in mobile sensing does not study the effect of the geographical diversity of users (e.g., country of residence) on smartphone sensing-based mood inference models.

Issues of generalization and fairness with regard to the geographical diversity of data sources have been discussed extensively in domains such as computer vision, speech, and natural language processing \cite{grother2019face, zou2018ai, wang2020towards, castelvecchi2020facial, malhotra2008automatic}. For example, gender classification models trained with data predominantly from the USA have performed poorly on people of African and Asian descent \cite{castelvecchi2020facial}. Many geographical-related biases (e.g., Indian brides being recognized as dancers, etc.) have been shown in models trained with the imagenet dataset, in which a majority of data is from western countries \cite{zou2018ai}. Such findings have uncovered issues in data collection practices and helped shape research directions to address issues related to diversity and biases. In this context, many prior mobile sensing studies that attempt inferences regarding well-being related aspects highlighted that models are trained in specific countries, and the generalization of techniques for other countries or regions should be explored further \cite{choi2021kairos, muller2021depression, meegahapola2021examining, meegahapola2021one}. However, mood inference studies have focused on only one or two countries for data collection \cite{likamwa2013moodscope} or have not considered the diversity of data sources in terms of the country, even when data were collected from multiple countries \cite{servia2017mobile}. 

Bardram et al. \cite{bardram2020decade} emphasized the need for generalization and reproducibility of sensing-based models for mental well-being-related outcomes. However, even though examining gender, age, and occupation-related diversity is feasible even within the same country, examining geographical diversity requires a considerable effort in conducting the same study, with the same protocol, in several geographic regions because studies are time-consuming and expensive; and logistical difficulties in conducting experiments such as language barriers, technology barriers, differences in motivating use cases and required incentives. Hence, studies that examine the geographical diversity of mobile sensing-based inferences are rare \cite{phan2022mobile, khwaja2019modeling}. In this paper, we study and compare the performance of country-specific, country-agnostic, and multi-country approaches for mood inference. In addition, we also examine the effects of model personalization and generalization to new geographically diverse countries. To our knowledge, this is one of the first studies to examine the effect of geographical diversity of users on smartphone sensing-based mood inference models, hence shedding light on distributional shift related issues. Considering these aspects, we ask three research questions. 

\begin{itemize}[wide, labelwidth=!, labelindent=0pt]
    \item[\textbf{RQ1:}] What behavioral and contextual characteristics around mood reports of college students (from eight countries spanning Europe, Asia, and Latin America) can be extracted from the analysis of smartphone sensing and self-report data?
    \item[\textbf{RQ2:}] How do smartphone sensing-based mood inference models perform in different countries (country-specific)? Can a model trained in one/more countries be deployed in another country not seen on training data to achieve reasonable accuracies, hence generalizing well (country-agnostic)?
    \item[\textbf{RQ3:}] How do country-specific or continent-specific models perform as compared to a multi-country model? 
\end{itemize}{}

By addressing the above research questions, this paper provides the following contributions: 
\begin{itemize}[wide, labelwidth=!, labelindent=0pt]
    \item[\textbf{Contribution 1:}] We conducted a new smartphone-based data collection campaign among 678 participants in eight countries (China, Denmark, India, Italy, Mexico, Mongolia, Paraguay, UK) representing Europe, Asia, and Latin America to study their everyday mood and behavior. During the study, we collected 329,974 fully complete self-reports. In addition, we also collected rich passive sensing data with continuous sensing (activity type, step count, location, cellular, wifi, bluetooth, proximity, etc.) and interaction sensing (app usage, touch events, user presence, screen-on/off episodes, notifications, etc.) throughout the deployment. First, we found that negative mood reports in all countries would increase from morning to night. Moreover, with statistical analysis, we found that the features that help infer mood are different across countries. However, the best features included both continuous and interaction sensing modalities in all countries. 
    
    \item[\textbf{Contribution 2:}] We found that the country-specific approach performs reasonably for both two-class and three-class mood inferences with AUROC scores in the range of 0.76-0.98 with hybrid (i.e., partially personalized) models. However, we noticed that across both two-class and three-class inferences, models do not generalize well to other countries, where AUROC scores drop to the range of 0.46-0.55 on average in the population-level (i.e., non-personalized) setting and 0.66-0.73 in the hybrid setting. These findings raise the significance of discussing issues of generalization of mobile sensing-based models to different world regions. 
    
    \item[\textbf{Contribution 3:}] In the hybrid setting, we found that multi-country models do not perform as well as country-specific models even though they achieved an AUROC of 0.81. However, they performed better than continent-specific models built for Asia and worse than the one built for Europe. Even though the performance differences were not high, this again highlights that building a model within European countries leads to higher performance and better generalization for those countries than using multi-country or even some country-specific models. A possible explanation is that the European countries under study (Italy, Denmark, UK) might share some daily behavioral patterns. In contrast, the three countries in Asia under study (China, India, Mongolia) have less similarity regarding daily patterns. Hence, these findings point toward the benefit of considering the geographical/cultural diversity of data collection on smartphone sensing-based mood inference models. 
\end{itemize}{}

The study is organized as follows. In Section~\ref{sec:related_work}, we describe the background and related work. Then we describe the data collection procedure in eight countries and how we came up with features in Section~\ref{sec:mobile_app}. Section~\ref{sec:data_analysis} provides a descriptive and a statistical analysis of data. In Section~\ref{sec:inference}, we define the analysis strategy and evaluate two-class and three-class mood inference with population-level and hybrid models with approaches: country-specific, continent-specific, country-agnostic, and multi-country. We discuss the main findings and implications in Section~\ref{sec:discussion}, and conclude the paper in Section~\ref{sec:conclusion}.

\section{Background and Related Work}\label{sec:related_work}

\begin{table*}
\centering
\caption{Terminology and description regarding different model types and approaches.}
\label{tab:terminology}
\resizebox{\textwidth}{!}{%
\begin{tabular}{>{\arraybackslash}m{2.1cm} >{\arraybackslash}m{15cm}}

\rowcolor{Gray2!20}
\textbf{\makecell[l]{Terminology}}  & 
\textbf{Description} 
\\

\arrayrulecolor{Gray}

Population-Level Model (PLM) &
Training and Testing splits have a disjoint set of users. Represents a case where a machine learning model trained with a population is deployed to a mobile app that is used by a new user. Hence, end user data are not used in model training leading to non-personalized and generic one-size-fits-all models.\\

\rowcolor{Gray!15}
Hybrid Model (HM)&
Training and testing splits do not have a disjoint set of users. Represent a case where a machine learning model is used by a mobile app user for some time, and data from the user is used in re-training models. Hence, this approach leads to partially personalized models. \\

\hline

\textcolor{black}{Country-Specific} &
\textcolor{black}{This approach uses training and testing data from the same country. 
Each country has its own model, without leveraging data from other countries. As the name indicates, these models are specific to each country (e.g., a model trained in Italy and tested in Italy). Both population-level and hybrid model types can be trained in the country-specific approach.}\\

\rowcolor{Gray!15}
\textcolor{black}{Continent-Specific} &
\textcolor{black}{This approach uses training and testing data from the same continent. Each continent has its own model, without leveraging data from other continents. As the name indicates, these models are specific to each continent (e.g., a model trained in Europe and tested in Europe). Continent specific approach can be trained with population-level and hybrid models.}\\

\textcolor{black}{Country-Agnostic} &
\textcolor{black}{This approach assumes that data and models are agnostic to the country. Hence, a trained model can be deployed to any geographical region regardless of the country of training. Country-agnostic approach too can be trained with population-level and hybrid models. There are two types of country-agnostic settings:\newline (1) Country-Agnostic I: The first setting uses training data from one country, and testing data from another country. This corresponds to the scenario where a model trained a in country already exists, and we need to understand how it would generalize to a new country (e.g. a model trained in Italy and tested in Mongolia). \newline (2) Country-Agnostic II: The second setting  uses training data from four countries, and testing data from the remaining country. This corresponds to a scenario where the model was already trained with data from several countries, and we need to understand how it would generalize to a new country (e.g. a model trained with data from Italy, Denmark, UK, and Paraguay, and tested in Mongolia).} \\

\rowcolor{Gray!15}
\textcolor{black}{Multi-Country}&
\textcolor{black}{This one-size-fits-all approach uses training data from all eight countries and tests the learned model in all countries. This corresponds to the setting in which multi-country data is aggregated to build a single model. However, this is also how models are typically built without considering aspects such as geographical diversity. Multi-Country models too can be trained with population-level and hybrid approaches.} \\

\arrayrulecolor{Gray2}
\hline

\end{tabular}
}
\end{table*} 

\subsection{Definitions and Terminology}

\subsubsection{What is Mood?} 

There is no single way to define mood \cite{dissanayake2022sigrep}. However, in prior work in mobile sensing, some operationalizations have been commonly used. Positive Negative Affect Schedule (PANAS) is a widely used validated questionnaire that can be used to capture the positive and negative affect of individuals \cite{kanjo2017notimind}. \textcolor{red}{In addition, the Patient Health Questionnaire (PHQ-9) has been used in the past to quantify depressive mood with mobile sensing \cite{wahle2016mobile}}. However, these questionnaires are long and could be cumbersome to users \cite{likamwa2013moodscope}. Further, they can capture mood over the past week (or two), \textcolor{red}{and might not be suitable to measure the in-situ mood for long time periods}. Hence, prior work has also used an affect grid based on the circumplex mood model \cite{servia2017mobile, likamwa2013moodscope} that would capture the \emph{valence} and \emph{arousal}. As described in later sections, due to pragmatic reasons, the data collection in this study does not focus on arousal because positive and negative affects of the circumplex model are important in determining negative moods that could be useful for adverse mental well-being related outcome detection, feedback, and interventions \cite{baumel2019objective,schueller2021understanding}. Hence, only \emph{valence} has been captured in a five-point scale: very positive (\includegraphics[scale=0.13]{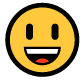}), positive (\includegraphics[scale=0.13]{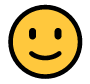}), neutral (\includegraphics[scale=0.13]{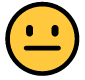}), negative (\includegraphics[scale=0.13]{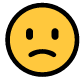}), very negative (\includegraphics[scale=0.13]{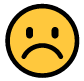}). This five-point scale is similar to LiKamwa et al. \cite{likamwa2013moodscope} and Horlings et al. \cite{horlings2008emotion}. For inference, we reduce the five-point scale to two-point and three-point scales similar to prior work \cite{wampfler2022affective, dissanayake2022sigrep, brown2011towards}. This is usually done based on the idea that in mood inference, the more important aspect is to detect extreme moods (i.e., negative, positive) rather than to identify all fine-grained intermediate mood levels in the middle of the spectrum \cite{horlings2008emotion}. First, obtaining a three-point scale using the five-point scale was obvious by combining very positive and positive to positive; neutral as it is; and negative and very negative to negative, hence having three classes \cite{wampfler2022affective, soleymani2011multimodal}. However, for two-class inference, the categorization is not as obvious. Some prior studies have removed the class in the middle (i.e., neutral), hence obtaining positive and negative labels \cite{zhang2020corrnet, horlings2008emotion}. Even though it is possible to do it with the available classes in the dataset, we believe it would lead to a biased classifier that would not perform reasonably well when exposed to data corresponding to neutral mood labels. Hence, we followed prior work that binned very positive, positive, and neutral moods as positive; and negative and very negative moods as negative \cite{zhang2020corrnet, brown2011towards}. This two-class inference also allows for detecting negative moods, which is useful in mobile health apps for feedback and interventions \cite{baumel2019objective,schueller2021understanding} because it is such negative moods, along with other aspects like stress that could be harmful to individuals on the long term. Hence, in the scope of this paper, mood can be defined as \emph{the instantaneous valence reported by study participants on a five-point scale (from very positive (\includegraphics[scale=0.13]{images/verypositive.PNG}) to very negative (\includegraphics[scale=0.13]{images/verynegative.PNG})), reduced to either a two-point scale corresponding to positive and negative classes or a three-point scale corresponding to positive, neutral, and negative classes, for inference using smartphone sensing data.}

\subsubsection{Model Types and Approaches.}

This section introduces the definitions and terminology used in this paper, as summarized in Table~\ref{tab:terminology}. In terms of model types, we use population-level (subject-independent) and hybrid models \cite{ferrari2020personalization, bangamuarachchi2022sensing, likamwa2013moodscope}. While population-level models are not personalized, hybrid models are partially personalized. The operationalization of models is described in Section~\ref{sec:inference}. Second, in terms of approaches, we consider the country-specific approach that is trained and tested within each country; the continent-specific approach that is trained and tested within each continent; the country-agnostic approach in which models are trained in one or more countries, and tested in an unseen country; and the multi-country approach that would ignore the diversity in terms of countries, and train a one-size-fits-all model considering data from all countries. As an important note, all these approaches can be evaluated with both population-level and hybrid model types. For example, in a country-specific setting, imagine a model trained with a certain population in Italy and tested with some new users in Italy, hence examining the model performance on new users from the same country. This is equivalent to a population-level model of the country-specific approach. Then, imagine the set of unseen users producing data for model training after using a mobile app for some time, and these data points being used to update the model. This would then lead to a hybrid model of the country-specific approach. Similarly, for the country-agnostic approach, a model trained in Italy deployed to unseen users in Paraguay is similar to evaluating a population-level model. Then, imagine the users in Paraguay providing some data for model personalization. This leads to a hybrid model created with a mix of data from Italy and Paraguay that can be evaluated on new data points from users in Paraguay, whose data were used in model training. While this model too can be called a multi-country model, for ease of understanding in the scope of this paper, we would still call it a hybrid model with the country-agnostic approach. Using the combination of model types and approaches, we can examine the effect of personalization (with model types) and model generalization to new countries (with the four approaches), hence uncovering distributional shift-related issues of multi-modal mobile sensing datasets for mood inference.

\subsection{Considerations for Research in Mobile Sensing Involving Geographic Diversity}

\subsubsection{Mood and Geographical Diversity} 

Across different geographical regions and cultures, behavior is mediated by inherent beliefs, presses, and affordances of physical and/or socio-cultural environments \cite{phan2022mobile}. Even for behaviors that are similar across cultures, the psychological meaning of those behaviors might not be the same due to \cite{phan2022mobile}: (a) Certain behaviors that are acceptable in certain countries/cultures are not perceived as normative or appropriate in other countries \cite{van2004bias}; (b) The same behavior might be indicative of different outcomes/functions. For example, while cycling is everyday behavior in certain regions (e.g., Aalborg, Denmark), it might only be used for exercise in other areas (e.g., Ulanbatoor, Mongolia); and (c) Different behaviors might be indicative of a similar outcome/function. For example, while people in some countries might perform cycling for exercise, people in other countries might prefer going to the gym for exercise. Why people cycle will depend on many contextual and cultural factors such as road safety, availability of public transport, alternative exercise options, weather conditions, and perceptions about cycling in a specific geographical region. Given that smartphone sensors can capture such physical activities (e.g., Google Activity Recognition API \cite{GoogleActivity2022} and other activity engines built by researchers \cite{wang2014studentlife}) and are used to infer more complex variables \cite{canzian2015trajectories, wang2014studentlife}, invariably, such behavioral differences across geographical areas could affect mood inference models that leverage activity data from accelerometers and location \cite{phan2022mobile}. In addition, device-mediated behavior or phone usage behavior could also vary between geographical areas depending on cultural norms, weather conditions (e.g., the phone usage behavior while walking outside in a cold vs. a hot country), network coverage, and subscription plans (e.g., people in countries where internet plans are expensive might turn off internet frequently, people in countries where the used phones are old might turn off Wifi and location sensors often to save battery of the phone, etc.), and availability of alternative equipment that could serve similar functionality (e.g., using a laptop for zoom calls instead of the phone, hence showing differences in the sensed app usage behavior). Given that mood inference models in prior work have used both continuous (activity types, step counts, location, proximity, wifi, etc.) and interaction (typing and touch events, user presence, application usage, screen on and off events, etc.) sensing modalities to examine/infer mood and related psychological constructs, how behaviors and contexts captured with smartphones affect mood inference in different countries is worth investigating. 

\subsubsection{Studies about Psychological Constructs and Geographical Diversity.} According to Khwaja et al. \cite{khwaja2019modeling}, psychological mobile sensing research aims to quantify and measure constructs related to mood, stress, depression, and user personality over the last decade due to the advancement of sensing technologies. Even though there is a myriad of studies about such psychological aspects, ranging from clinical to non-clinical studies, many have focused on a population within a single country \cite{phan2022mobile}. In addition, even when the construct of analysis used in studies is the same (e.g., circumplex mood model, positive-negative affect schedule, etc.), comparing different studies across countries is complicated because data have been collected using different protocols and sensing modalities \cite{adler2022machine}. Furthermore, Phan et al. \cite{phan2022mobile} have discussed how prior psychology studies in mobile sensing have collected data focusing on WEIRD samples (Western, Educated, Industrialized, Rich, and Democratic) and paid less attention to the global south. This has also been highlighted in a review study on smartphone sensing by Meegahapola et al. \cite{meegahapola2020alone}. For these reasons, prior work has emphasized the need for studies that examine the generalization of models across countries/cultures by building diversity-aware approaches to machine learning-based modeling of sensor data \cite{meegahapola2020smartphone, bardram2020decade}. According to a recent review by Phan et al. \cite{phan2022mobile}, only Khwaja et al. \cite{khwaja2019modeling} have considered the cultural diversity of smartphone sensing-based models on psychological aspects, where they studied personality traits based on Big-Five model. In that study, the authors collected data from 166 participants from five countries (UK, Spain, Colombia, Peru, and Chile). They showed that country-specific models perform the best, regardless of the gender or age balance, for the prediction of Extraversion, Agreeableness, and Conscientiousness. Compared to that study, we also collected data from multiple countries. However, our primary focus is on studying mood inference models that could vary from time to time, even within the same person (more dynamic), instead of stable personality traits. In addition, Muller et al. \cite{muller2021depression} used mobile GPS data to predict depression in socio-demographically homogeneous sub-samples within the USA. They trained algorithms for the whole sample and homogeneous sub-samples (e.g., highly educated men, women residing in rural regions, etc.) and tested within and across sub-samples. They found that the technique that led to high AUROC scores for student populations (0.82), did not generalize well to the USA-wide population-level model (AUROC of 0.57). In contrast, our work focuses on valence instead of depressive mood. In addition, rather than concentrating on socio-demographic differences within a particular country, we focus on cross-country differences.  

\subsection{Mood and Smartphone Technologies}

\subsubsection{Mood Tracking with Self-Reports} 

In the early days, mobile phone-based mood charts were used to track the mood of individuals. These are based on self-reported questionnaires and ecological momentary assessment (EMA) responses \cite{chan2018asynchronous, meegahapola2020smartphone}. Similar to how mobile food diaries were designed for people who wanted to control their diet \cite{meegahapola2021one}, mood charts were designed to support people who wanted to control negative moods and increase self-awareness, allowing for monitoring and feedback \cite{baumel2019objective,schueller2021understanding}. With randomized controlled trials, some studies explored the usefulness and efficacy of self-report-based mood tracking and showed that engaging in mood tracking tools increases self-awareness, hence reducing the possibility of having anxiety, even within clinically depressed populations \cite{bakker2018moodprism, birney2016moodhacker}. Going beyond applications related to health and well-being, Glasgow et al. discussed how aspects like destinations, travel choices, and social ambiance are related to mood \cite{glasgow2019travel}. Further, in this context, prior work that uses mood tracking has focused on different populations such as college students \cite{lee2018destressify, wang2014studentlife}, adolescents \cite{kenny2015copesmart} and clinically diagnosed, high-risk populations with mental well-being related issues \cite{wang2016crosscheck, faherty2017pregnancy, mark2008chart}. Hence, most prior studies relied on user engagement to keep track of mood. This could be a burden to users in the long run, and it is known that apps that require many self-reports do not have high adoption rates. In our work, even though we captured self-reports about mood, they were captured as ground-truth labels to train classifiers with sensor data for mood inference. Such inferences could be used to update mood-tracking applications that could be used to provide context-aware interventions, and feedback to users, with less user burden \cite{servia2017mobile}.

\subsubsection{Mood Tracking with Sensing.} 

Mobile phone sensors allowed researchers to build context-aware systems that could infer various aspects regarding the health and well-being of people \cite{lane2010survey}. Most of such studies rely on using features captured from sensors in smartphones as proxies to personal attributes (mood, stress, etc.), behavior (eating, drinking, running, walking, etc.), and context (social context, semantic location, ambiance, etc.) \cite{meegahapola2020smartphone}. Hence, there are studies that infer aspects like mood \cite{servia2017mobile, likamwa2013moodscope}, stress \cite{lu2012stresssense, sano2013stress}, depression \cite{canzian2015trajectories, farhan2016behavior}, eating behavior \cite{meegahapola2021one,biel2018bites}, drinking behavior \cite{santani2018drinksense}, activity types \cite{morales2017physical}, and social contexts \cite{meegahapola2021examining, meegahapola2020alone}, among many others. If we specifically focus on mood-related studies, LiKamwa et al. \cite{likamwa2013moodscope} showed that the mood of individuals captured with the circumplex mood model could be inferred with an accuracy of 66\% with all user models (population-level), which can be increased up to 93\% using personalization (user-level) with a dataset collected from 32 individuals. They suggested that building hybrid models (partially personalized) would help overcome the drawbacks of both population-level and user-level models. Servia-Rodríguez et al. \cite{servia2017mobile} collected a large-scale dataset of mood self-reports and passive sensing data from multiple countries. They also showed that binary mood captured with the circumplex mood model could be inferred with an accuracy of 70\% with population-level models. Some studies examined mood instability derived using mood reports, with phone sensor data \cite{morshed2019prediction, zhang2019inferring}. In our work, we look into inferring mood valence with population-level and hybrid models. However, we are more interested in examining (a) the similarities and differences in mood models for different countries; and (b) the generalization of models to unseen countries, both of which have not been examined in prior work. Further, as Bardram et al. \cite{bardram2020decade} highlighted, there is a lack of reproducibility and generalization of machine learning models across studies in this domain. We believe the results presented in this study would be a step in the right direction for better awareness of these issues in examining the characteristics and generalization of smartphone sensing-based mood inference models across different geographical regions.

\section{Study Design, Data Collection, and Feature Extraction}\label{sec:mobile_app}

\begin{figure}
    \includegraphics[width=\textwidth]{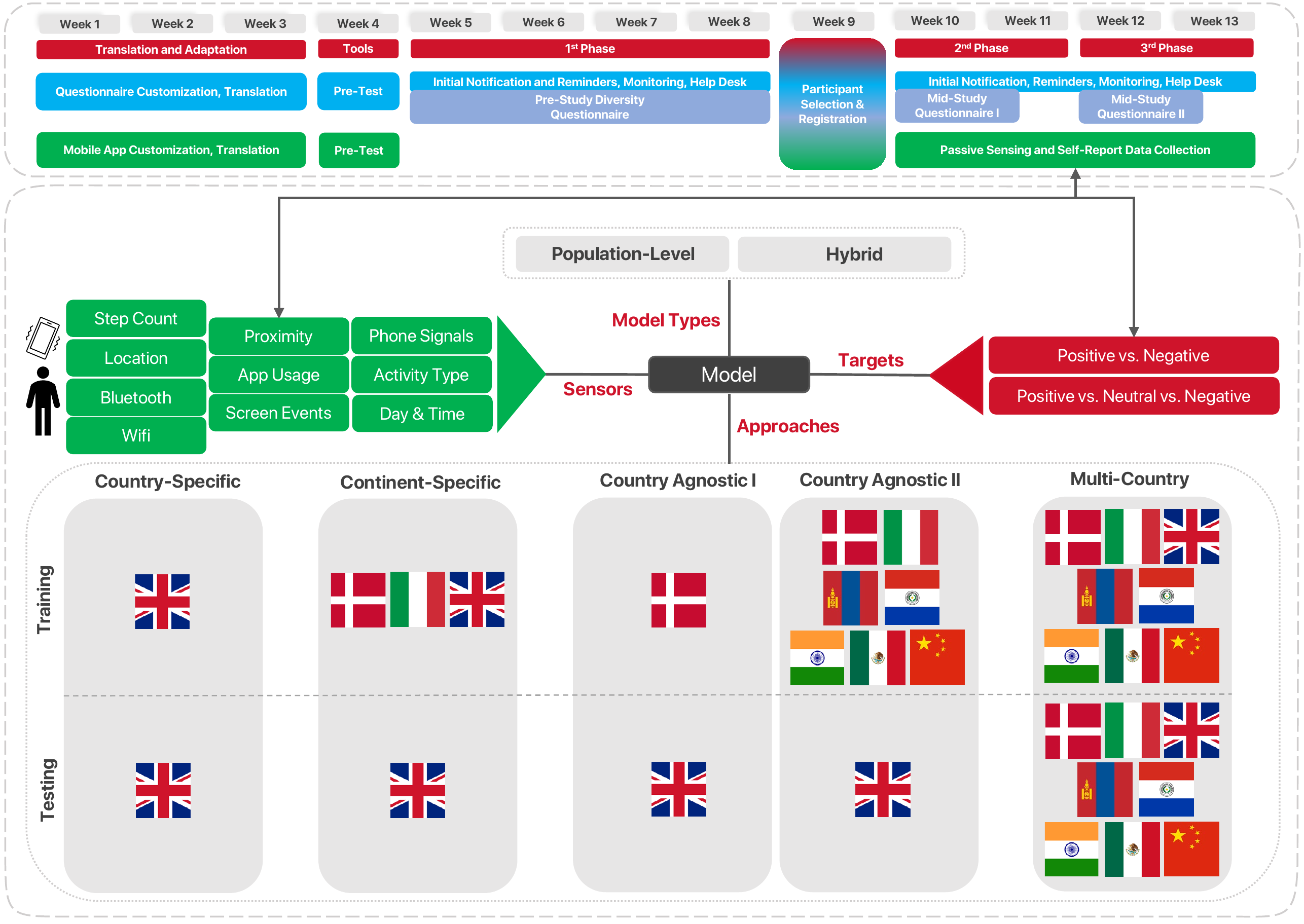}
    \caption{High-level overview of the study.}
    \label{fig:overview}
\end{figure}

With a team representing computer science, social sciences, user experience design, and ethics from institutions in over ten countries, we designed an exploratory study and developed a mobile sensing application to collect passive smartphone sensing and self-report data from participants about their everyday life behavior and well-being. The ultimate goal of this deployment was to study their behavior, including aspects such as activity, social context, location, mood, and sleep quality from a mobile sensing standpoint and also to consider various diversity aspects that could potentially affect sensing-based inferences (ranging from geographical region and gender to personality and values). The study is summarized in Figure~\ref{fig:overview}. The study design consisted of two main components: (a) LimeSurvey component to collect survey responses during pre and mid-study phases; and (b) Mobile sensing app to collect sensor data and self-reports. A technical report regarding the study procedure and future plans for dataset access is available in \cite{giunchiglia2022aworldwide}.

\subsection{Survey Questionnaires}

Survey responses were captured from participants with three questionnaires sent to them before and during the pilot at three different times. This was done to ensure that the burden on participants was reasonable. These questionnaires were administered through the LimeSurvey platform \cite{engard2009limesurvey}. 

\subsubsection{Pre-Study Diversity Questionnaire}
The primary objective of this questionnaire was to capture diversity attributes of participants from different perspectives. As the first step, basic demographic information was captured, including gender, age, sex, degree program, and socioeconomic status. Then, in an attempt to capture the psychosocial profile, the 20-item Big Five Inventory (BFI) \cite{donnellan2006mini} and Basic Value Survey (BVS) \cite{gouveia2014functional} were administered. Finally, there were several questions regarding social relationships (virtual and real) and cultural consumption that they were interested in.

\subsubsection{Mid-Study Questionnaire I and II}
The objective of the first questionnaire was to gather more detailed information about personality using the Jungian Scale on Personality Types \cite{jung2016psychological} and Human Values Survey \cite{schwartz1994there}. In addition, questions regarding physical activity and sports, cooking and shopping habits, transport methodologies, and cultural activities were captured. The second questionnaire consisted of the Multiple Intelligences Profiling Questionnaire \cite{tirri2008identification}. This also contained several open-ended questionnaires about the mobile app user experience. 

\subsection{Mobile Application}\label{subsec:mobile_app}

An Android mobile application was used to capture the everyday behavior of participants using short in-situ self-report questions. The app was developed such that data would be stored in an SQLite database on the phone, and later when the phone is connected to a Wifi network, data would be uploaded to the main server and free up the local phone storage. In addition, the app could send push notifications by using Google Firebase as a notification broker. Hence, the three main components of the application are: (1) a push notification system that would send periodical reminders to participants to fill in self-reports; (2) mobile time diaries to capture self-reports; and (3) a smartphone sensing component to collect passive sensing data from multiple modalities.

\subsubsection{Push Notifications.} Given the nature of the study and the requirement to capture behavioral and situational data in a particular moment, the app sent reminders for participants to fill in in-situ self-reports regarding their everyday life behavior around 20 times throughout the day. In addition, start-of-the-day and end-of-the-day questionnaires were administered at the beginning and end of the day. When a notification was not clicked and a participant did not complete the self-report within two hours, the notification expired, and a new notification would be sent later. This allowed to keep track of participant compliance (e.g., how many self-reports were answered from the total number of notifications sent). 

\subsubsection{Time Diaries and Start/End-of-the-Day Questionnaires.} The start-of-the-day questionnaire was sent to participants at 8 am each day. It only had two questions with five-point likert scales (very good to very bad): (i) sleep quality; and (ii) expectations about the day. The end-of-the-day questionnaire was sent to participants at 10 pm and asked them (a) to rate their day (five-point likert scale; very good to very bad); (b) if they had any problems during the day (open response), and (c) how did they solve them (open response). The time diary was sent to users once every 30-60 minutes. While this allowed capturing longitudinal behavior granularly, it also introduced user burden. Therefore, the time diary was designed to minimize user burden and reduce completion time. Hence, after several iterations of discussions, only four questions were included in this component: (i) current activity: 34 activities including eating, working, attending a lecture, etc.; (ii) semantic location: 26 categories including home, workplace, university, restaurant, etc.; (iii) social context: 8 categories including alone, with the partner, family member/s, friends, etc.; and (iv) current mood: five-point likert scale to capture the valence of the circumplex mood model \cite{russell1980circumplex} similar to LiKamwa et al. \cite{likamwa2013moodscope}, with an emoji-scale. As explained in Section~\ref{sec:related_work}, this is the variable we chose as this paper's primary focus.

\begin{table*}
    \caption{\textcolor{red}{Summary of 105 features extracted from sensing data, aggregated around activity self-reports using a time window. A detailed description about sensing modalities is provided in Appendix A.}}
    \label{tab:agg-features}
    \begin{tabular}{{p{0.139\linewidth}p{0.11\linewidth}p{0.725\linewidth}}}
        
        \cellcolor[HTML]{EDEDED}\textbf{Modality} &
        \cellcolor[HTML]{EDEDED}\textbf{Frequency} &
        \cellcolor[HTML]{EDEDED} \textbf{Features and Description}
        \\

        \makecell[l]{Location} & 
        \makecell[l]{1 sample \\per minute} & 
        \makecell[l]{radius of gyration, distance traveled, mean altitude} \\
\arrayrulecolor{Gray}
    \midrule
    
        \makecell[l]{Bluetooth \\{[}low energy, \\normal{]}} &
        \makecell[l]{1 sample \\per minute} & 
        \makecell[l]{number of devices (the total number of unique devices found), mean/std/min/max \\rssi (Received Signal Strength Indication -- measures how close/distant other \\devices are)} \\

\arrayrulecolor{Gray}
    \midrule
    
        \makecell[l]{WiFi} & 
        \makecell[l]{1 sample \\per minute} & 
        \makecell[l]{connected to a network indicator, number of devices (the total number of unique \\devices found), mean/std/min/max rssi} \\
        
\arrayrulecolor{Gray}
    \midrule
    
        \makecell[l]{Cellular {[}GSM, \\WCDMA, LTE{]}} & 
        \makecell[l]{1 sample \\per minute} & 
        \makecell[l]{number of devices (the total number of unique devices found), mean/std/min/max \\phone signal strength} \\
        
\arrayrulecolor{Gray}
    \midrule
    
        \makecell[l]{Notifications} & 
        \makecell[l]{on change} & 
        \makecell[l]{notifications posted (the number of notifications that came to the phone), \\notifications removed (the number of notifications that were removed by the \\user) -- these features were calculated with and without duplicates.} \\
        
\arrayrulecolor{Gray}
    \midrule
    
        \makecell[l]{Proximity} & 
        \makecell[l]{10 samples \\per second} & 
        \makecell[l]{mean/std/min/max of proximity values} \\
        
\arrayrulecolor{Gray}
    \midrule
        \makecell[l]{Activity} & 
        \makecell[l]{2 samples \\per minute} & 
        \makecell[l]{time spent doing activities: still, in\_vehicle, on\_bicycle, on\_foot, running, tilting, \\walking, other (derived using the Google activity recognition API \cite{GoogleActivity2022})} \\
        
\arrayrulecolor{Gray}
    \midrule
        \makecell[l]{Steps} & 
        \makecell[l]{10 samples \\per second \\ or on change} & 
        \makecell[l]{steps counter (steps derived using the total steps since the last phone turned on \\at 10 samples per second), steps detected (steps derived using event triggered for \\each new step captured on change)} \\
        
\arrayrulecolor{Gray}
    \midrule
        \makecell[l]{Screen events} &
        \makecell[l]{on change} & 
        \makecell[l]{number of episodes (episode is from turning the screen of the phone on until the \\screen is turned off), mean/min/max/std episode time (a time window could have \\multiple episodes), total time (total screen on time within the time window)} \\
        
\arrayrulecolor{Gray}
    \midrule
        \makecell[l]{User presence} &
        \makecell[l]{on change} & 
        \makecell[l]{time the user is present using the phone (derived using android API that indicate \\whether a person is using the phone or not)} \\
        
\arrayrulecolor{Gray}
    \midrule
        \makecell[l]{Touch events} &
        \makecell[l]{on change} & 
        \makecell[l]{touch events (number of phone touch events)} \\
        
\arrayrulecolor{Gray}
    \midrule
        \makecell[l]{App events} & 
        \makecell[l]{10 samples \\per minute} & 
        \makecell[l]{time spent on apps of each category derived from Google Play Store \cite{likamwa2013moodscope, santani2018drinksense}: \\action, adventure, arcade, art \& design, auto \& vehicles, beauty, board, books \& \\
        reference, business, card, casino, casual, comics, communication,
        dating,\\education, entertainment, finance, food \& drink, health \&
        fitness, house, lifestyle,\\maps \& navigation, medical, music, news \&
        magazine, parenting, personalization, \\photography, productivity,
        puzzle, racing, role playing, shopping, simulation, \\social, sports,
        strategy, tools, travel, trivia, video players \& editors, weather, word, \\ not\_found}   \\

        \hline
    \end{tabular}
\end{table*}

\subsubsection{Sensor Data and Features} The app collected sensor data from a range of sensors passively. Hence, sensor data included continuous sensing modalities such as accelerometer, gyroscope, ambient light, location, magnetic field, pressure, activity labels generated by the google activity recognition API, step count, proximity, and available Wi-Fi and bluetooth devices. Interaction sensing modalities included application usage, typing and touch events, on/off screen events, user presence, and battery charging events. The modalities and features crafted from each modality are summarized in Table~\ref{tab:agg-features}. In feature engineering, interpretability was a key factor as all the features were defined in a meaningful manner. Similar to prior work in ubicomp, we used a time window-based approach for matching sensor data to self-reports \cite{servia2017mobile, likamwa2013moodscope, meegahapola2021one}. While different time windows can be chosen based on the application scenario, this paper presents results with a dataset created using a time window of 10 minutes. Hence, if the self-report regarding mood occurred at time $T$, sensor data would be considered from $T-5$ minutes to $T+5$ minutes. However, we also considered other time windows, such as 2, 4, 15, and 20 minutes. Results showed that the 10-minute time window performed better for this task. This could be because shorter time windows do not capture enough behaviors and contexts around self-reports to make a meaningful prediction regarding mood. Prior work has shown that larger time windows can capture a high amount of information about user behaviors \cite{bae2017detecting}. However, we can not use very large windows above 20 minutes because it would lead to a situation where sensor data segments for self-reports might get overlapped, leading to data overlap between samples. Therefore, throughout the paper, we present results with a ten-minute time window. In addition, in this paper, we do not discuss why each sensing modality was chosen and how features were derived. This is because such details have been discussed extensively in many prior studies on mobile sensing for health and well-being \cite{servia2017mobile, khwaja2019modeling, santani2018drinksense, likamwa2013moodscope, meegahapola2021one, bae2017detecting, biel2018bites, wang2014studentlife, meegahapola2020smartphone}.

\subsection{Participant Recruitment and Deployment}

The primary objective of this study was to capture data from diverse student populations. While many facets of diversity could be captured by experimenting within the same country, it is difficult to study geographical diversity in such a way. Hence, we conducted mobile sensing experiments in eight countries representing Europe, Asia, and Latin America. Details regarding the data collection are mentioned in Table~\ref{tab:num-participants}. According to prior work in mobile sensing, many studies have focused on Europe and North America, but not much research has been conducted in other world regions \cite{meegahapola2020smartphone, phan2022mobile}. Hence, conducting the same study with the same protocol in multiple countries allows to study mood inference models and geographical diversity in a novel sense. The study was conducted in the following phases.

\subsubsection{Translation and Adaptation} In this phase, each site received the English version of the questionnaires and the app, including time diaries and the list of sensors to be collected. These tools were evaluated and adapted, in coordination with all the partners, to the specific context (e.g., invitation letters, type and amount of incentives for the participants of the mobile app, privacy and ethics documentation, etc.). Some countries made minimal changes to better adapt the questionnaire to the local situation or academic organization. Concerning the standard scales mentioned above, the translations were completed by a forward translator from the original English version and then validated back via panel and back-translation processes by independent translators. In addition, whenever a validated questionnaire translation was available, we used it (e.g., the Big five traits questionnaire is readily available in several languages). After translation and adaptation, the tools were tested locally. A first test was conducted to check and validate the translations and evaluate the tools' usability. A second test was conducted by sending the questionnaires to a small sample of participants, both project partners and students from various universities. As far as questionnaires were concerned, approximately 30 participants were involved. This test was also used to ascertain the completion times. Concerning the mobile app, a two-week validation test was carried out.

\subsubsection{Invitations, Pre-Study Diversity Questionnaire, and Participants} This was the first of the three phases of the data collection. This phase started by sending an email containing the survey description, the invitation to the first questionnaire, and information on the second part of the data collection (sensing component) via university mailing lists. This invitation was then reiterated through four weekly reminders to all students who still needed to complete the survey. Over 20000 college students were contacted with mailing lists in the initial recruitment phase. Out of the set of people who were contacted, 13398 participants filled in the pre-study diversity questionnaire. Then, a subset of the eligible participants was selected to participate in the second part of the study, which was done with the mobile app. The requirements for the selection were two-fold: (i) having consented to the processing of personal data -- this required participants agreeing to release mobile data collected during the study after anonymization; and (ii) owning an Android smartphone compatible with the app.

\subsubsection{Mid-Study Questionnaire I, II and Mobile Sensing app} Of all the participants who completed the pre-study diversity questionnaire, 678 participants were chosen for the next phase with the mobile sensing app. This deployment was done between September and November 2020. The average age of study participants was 24.2 years (SD: 4.2), and the cohort had 58\% females. They were sent emails with a specification manual to download and install the mobile sensing app. In addition, the participants completed the mid-study questionnaire I. Reminders were sent after one week for participants who still needed to complete the questionnaire. Then, participants completed time diaries, and sensing data were passively collected in the mobile app. After two weeks of mobile sensing app usage, the mid-study questionnaire II was sent to participants via email. After sending out this questionnaire, two more weeks of mobile sensing data collection were conducted. Daily reports were produced to facilitate monitoring the time diary survey and identify possible problems, including: (1) the number of notifications each participant responded to; and (2) the amount of data collected by the individual sensors. Using this information, local field supervisors could contact the inactive participants every three days and support them as needed. A further element of contact was the daily sending of the results of a daily prize, which was an additional incentive for participants. Finally, this paper will only focus on the mood variable captured during the study, and deeper analyses around other questionnaires captured with pre-study, mid-study I, and mid-study II questionnaires will be done in future publications with different scopes.

\begin{table}
    \caption{Participants of the mobile sensing data collection (countries named in alphabetical order).}
    \label{tab:num-participants}
    \centering
    \begin{tabular}{llrrrr}

        \cellcolor[HTML]{EDEDED}\textbf{Country} &
        \cellcolor[HTML]{EDEDED}\textbf{University} &
        \cellcolor[HTML]{EDEDED}\textbf{Participants} &
        \cellcolor[HTML]{EDEDED}\textbf{$\mu$ Age ($\sigma$) } &
        \cellcolor[HTML]{EDEDED}\textbf{\% Women} &
        \cellcolor[HTML]{EDEDED} \textbf{\# Self-Reports }
        \\

        China &
        Jilin University &
        41 &
        26.2 (4.2) & 
        51 &
        22,289
        \\
        
        \arrayrulecolor{Gray}
        \hline
        
        Denmark &
        Aalborg University &
        24 &
        30.2 (6.3) & 
        58 &
        10,010
        \\
        
        \arrayrulecolor{Gray}
        \hline
        
        India &
        Amrita Vishwa Vidyapeetham &
        39 &
        23.7 (3.2) & 
        53 &
        4,233
        \\
        
        \arrayrulecolor{Gray}
        \hline
        
        Italy &
        University of Trento &
        240 &
        24.1 (3.3) &
        58 &
        151,342
        \\

        \arrayrulecolor{Gray}
        \hline
        
        Mexico &
        \makecell[l]{Instituto Potosino de Investigación\\ Científica y Tecnológica} &
        20 &
        24.1 (5.3) & 
        55 &
        11,662
        \\
        
        \arrayrulecolor{Gray}
        \hline
        
        Mongolia &
        National University of Mongolia &
        214 &
        22.0 (3.1) &
        65 &
        94,006
        \\
        
        \arrayrulecolor{Gray}
        \hline
        
        Paraguay &
         \makecell[l]{Universidad Católica \\"Nuestra Señora de la Asunción"} &
        28 &
        25.3 (5.1) &
        60 &
        9,744
        \\
        
        \arrayrulecolor{Gray}
        \hline
        
        UK &
        \makecell[l]{London School of Economics\\ \& Political Science} &
        72 &
        26.6 (5.0) & 
        66 &
        26,688
        \\

        \cellcolor[HTML]{EDEDED}\textbf{Total/Mean} &
        \cellcolor[HTML]{EDEDED}\textbf{} &
        \cellcolor[HTML]{EDEDED}\textbf{678} &
        \cellcolor[HTML]{EDEDED}\textbf{24.2 (4.2)} &
        \cellcolor[HTML]{EDEDED}\textbf{58} &
        \cellcolor[HTML]{EDEDED}\textbf{329,974} 
        \\
        
    \end{tabular}
    
\end{table}

\subsubsection{Incentive Design} An incentive scheme was designed to motivate participants to complete time diaries and provide sensing data. Incentives included monetary prizes for participants who completed at-least 85\% of time diaries (e.g., 20 Euro in Italy, 150 Kr in Denmark, etc.), cash prizes for multiple daily winners randomly chosen from each pilot (e.g., five winners were given a prize of 5 Euro in Italy, 5 MNT in Mongolia, etc.). In the end, three winners from each country were randomly chosen for a larger prize (e.g., 150 Euros per person in Italy, 150 Sterling Pounds in the UK, etc.). Incentives in all countries were designed by considering each country's socioeconomic status and expecting all participants to be compensated and motivated equally. 

\subsubsection{Ethical Procedures} All the survey activities and results at each site comply with the national ethical privacy-protecting laws and guidelines, hence getting approvals from respective ethical review boards. In addition, all the experiments, including non-European pilots, were compliant with the General Data Protection Regulation (GDPR) \cite{voigt2017eu}. Additionally, for non-European experiments, the activities and results have been developed to comply with those of a European country for compliance purposes. More specifically, Italian legislation was selected as the reference.

\section{Behavioral and Contextual Characteristics Around Mood Reports Extracted from Sensor Data and Self-Reports  (RQ1)}\label{sec:data_analysis}

\subsection{{Descriptive Analysis.}}\label{sec:descriptive_analysis}

\begin{figure*}[t]
\begin{center}

    \begin{subfigure}[t]{0.49\textwidth}
        \centering
        \includegraphics[width=\textwidth]{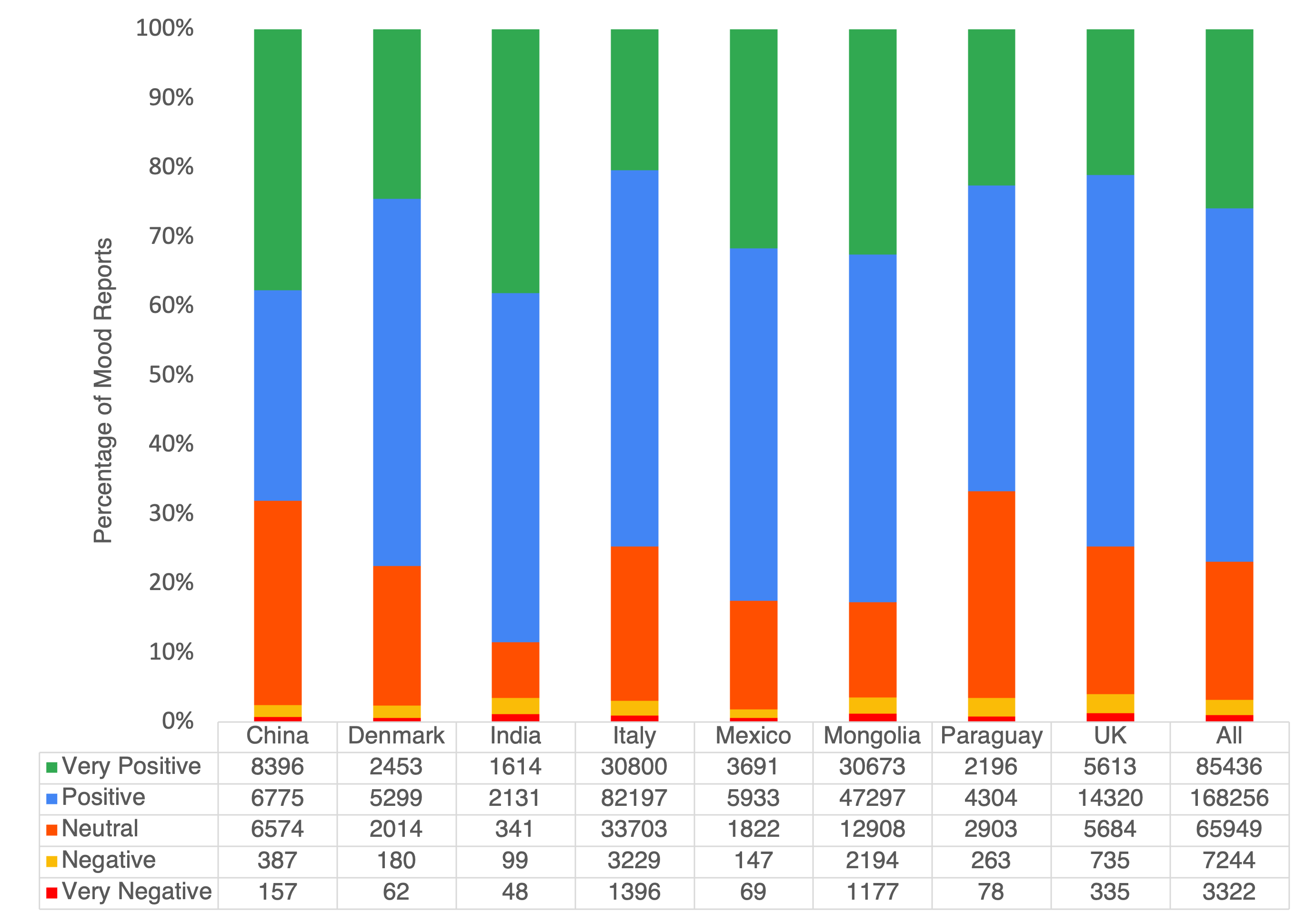}
        \caption{Five-class mood distribution}
        \label{fig:five_class_moods}
    \end{subfigure}
    \hfill 
    \begin{subfigure}[t]{0.49\textwidth}
        \centering
        \includegraphics[width=\textwidth]{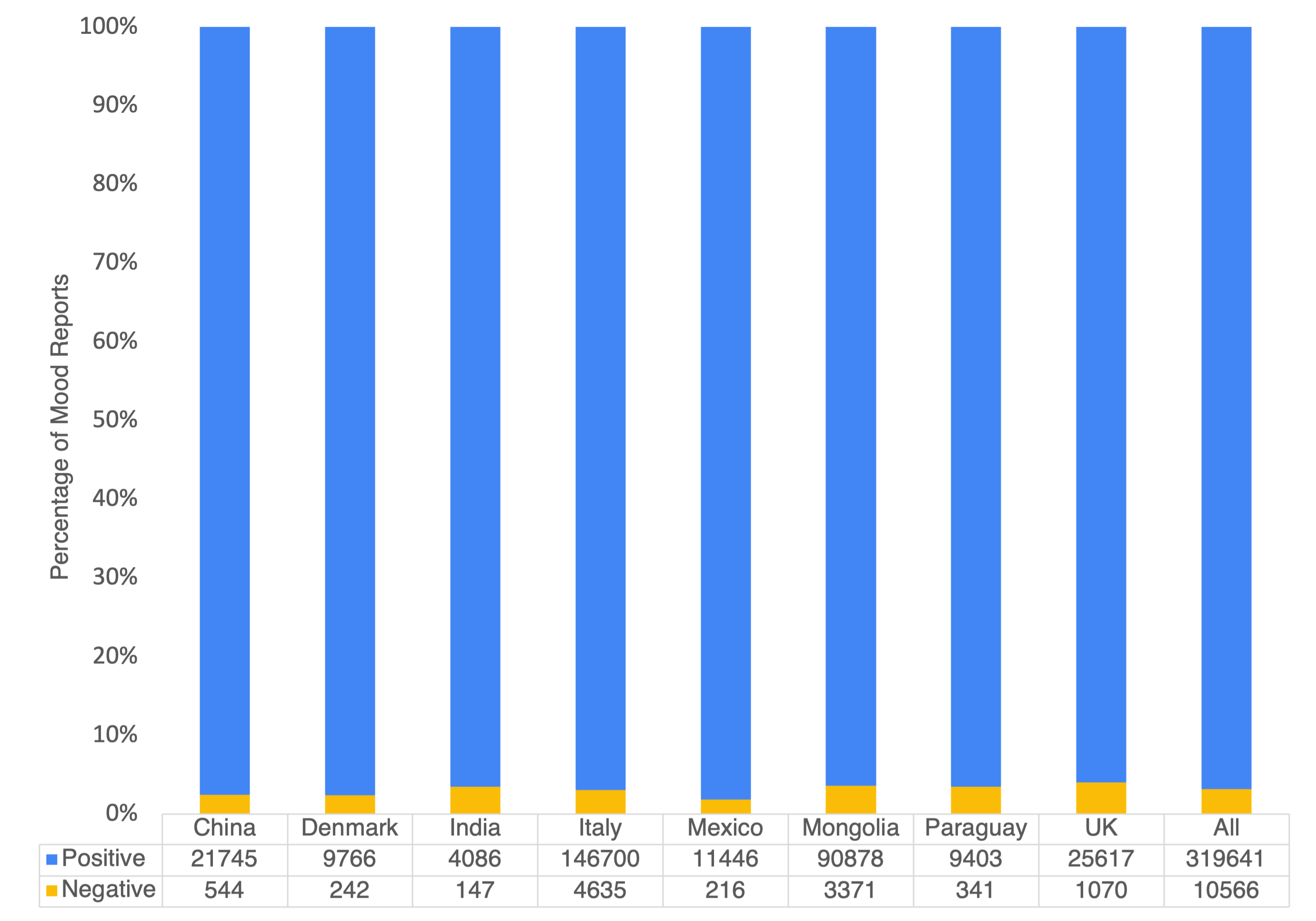}
        \caption{Two-class mood distribution}
        \label{fig:two_class_moods}
    \end{subfigure}
    \caption{Summary of self-reported mood distributions.}
    \label{fig:mood_distributions}
\end{center}
\vspace{-0.25 in}
\end{figure*}

\begin{figure*}[t]
\begin{center}

    \begin{subfigure}[t]{0.33\textwidth}
        \centering
        \includegraphics[width=1.125\textwidth]{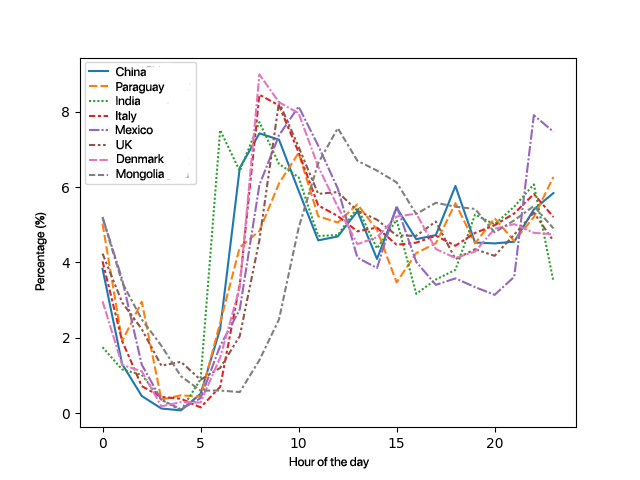}
        \caption{All}
        \label{fig:hourly_all}
    \end{subfigure}
    \hfill 
    \begin{subfigure}[t]{0.33\textwidth}
        \centering
        \includegraphics[width=1.125\textwidth]{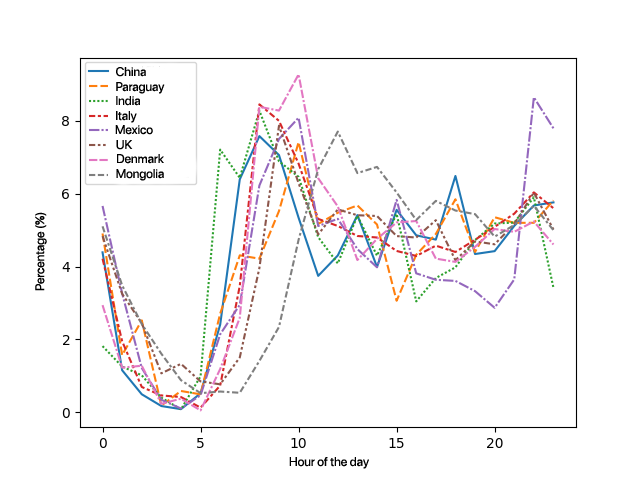}
        \caption{Positive}
        \label{fig:hourly_positive}
    \end{subfigure}
    \hfill 
    \begin{subfigure}[t]{0.33\textwidth}
        \centering
        \includegraphics[width=1.125\textwidth]{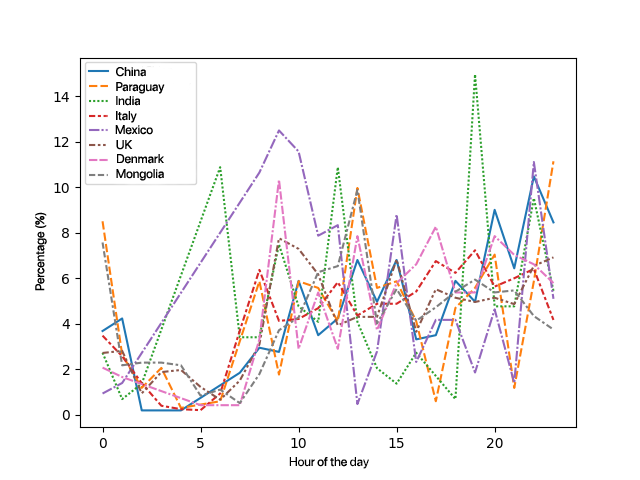}
        \caption{Negative}
        \label{fig:hourly_negative}
    \end{subfigure}
    
    \caption{\textcolor{red}{Distribution of self-reported moods for 24 hours of the day.}}
    \label{fig:hourly_distributions}
\end{center}
\vspace{-0.2 in}
\end{figure*}

Figure~\ref{fig:mood_distributions} shows the distribution of mood labels for the eight countries. We observed fewer labels for the ‘negative’ and ‘very negative’ classes compared to the ‘neutral’, ‘positive’, and ‘very positive’ classes. As shown in Figure~\ref{fig:five_class_moods}, except for China, where there were more ‘very positive’ reports than ‘positive’ or ‘neutral’ reports, all other countries had ‘positive’ as the majority label. This behavior of skewed reporting is common in studies about valence \cite{servia2017mobile, likamwa2013moodscope}. Furthermore, we plot the hourly distribution of mood reports in Figure~\ref{fig:hourly_distributions}. According to Figure~\ref{fig:hourly_all}, across all countries, we could see more self-reports in the morning compared to the afternoon or evening. However, Figure~\ref{fig:hourly_positive} shows that most self-reports around morning are for the positive valence. This means that users had a more positive mood after waking up and around the morning.
Interestingly, we also observed that the curve for Mongolia indicates late sleep and late wake-up, according to reports, which the partner institution later confirmed to be consistent with the routines of students in the country. As shown in Figure~\ref{fig:hourly_negative}, we also noticed that negative valence reports increase with time in most countries. This is in line with prior studies about mood and stress levels increasing with the time of the day \cite{mislove2010pulse}. 

\begin{figure*}[t]
\begin{center}

    \begin{subfigure}[t]{0.24\textwidth}
        \centering
        \includegraphics[width=1.125\textwidth]{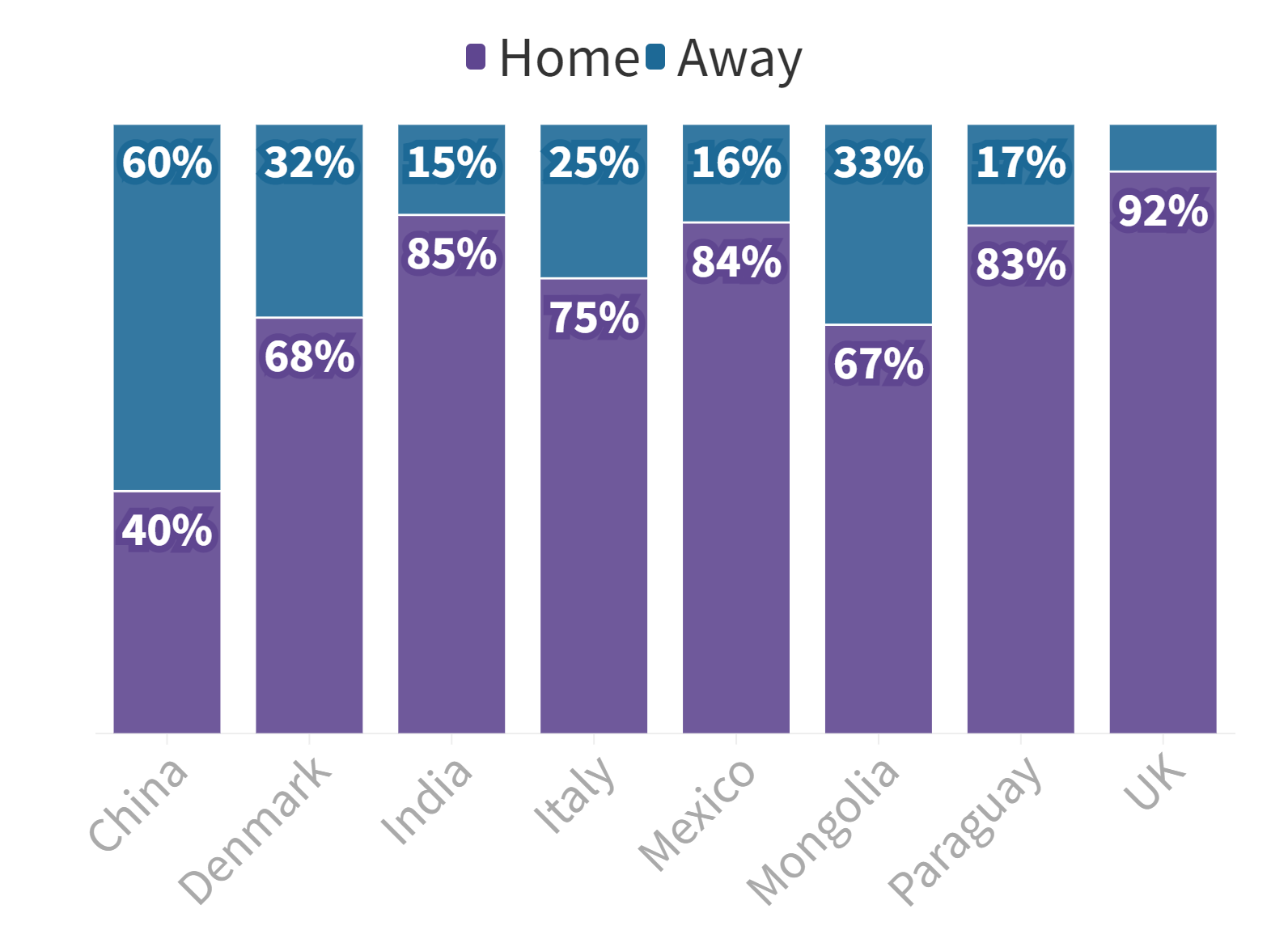}
        \caption{\centering Location context for negative mood.}
        \label{fig:neg_vs_lc}
    \end{subfigure}
    \hfill 
    \begin{subfigure}[t]{0.24\textwidth}
        \centering
        \includegraphics[width=1.125\textwidth]{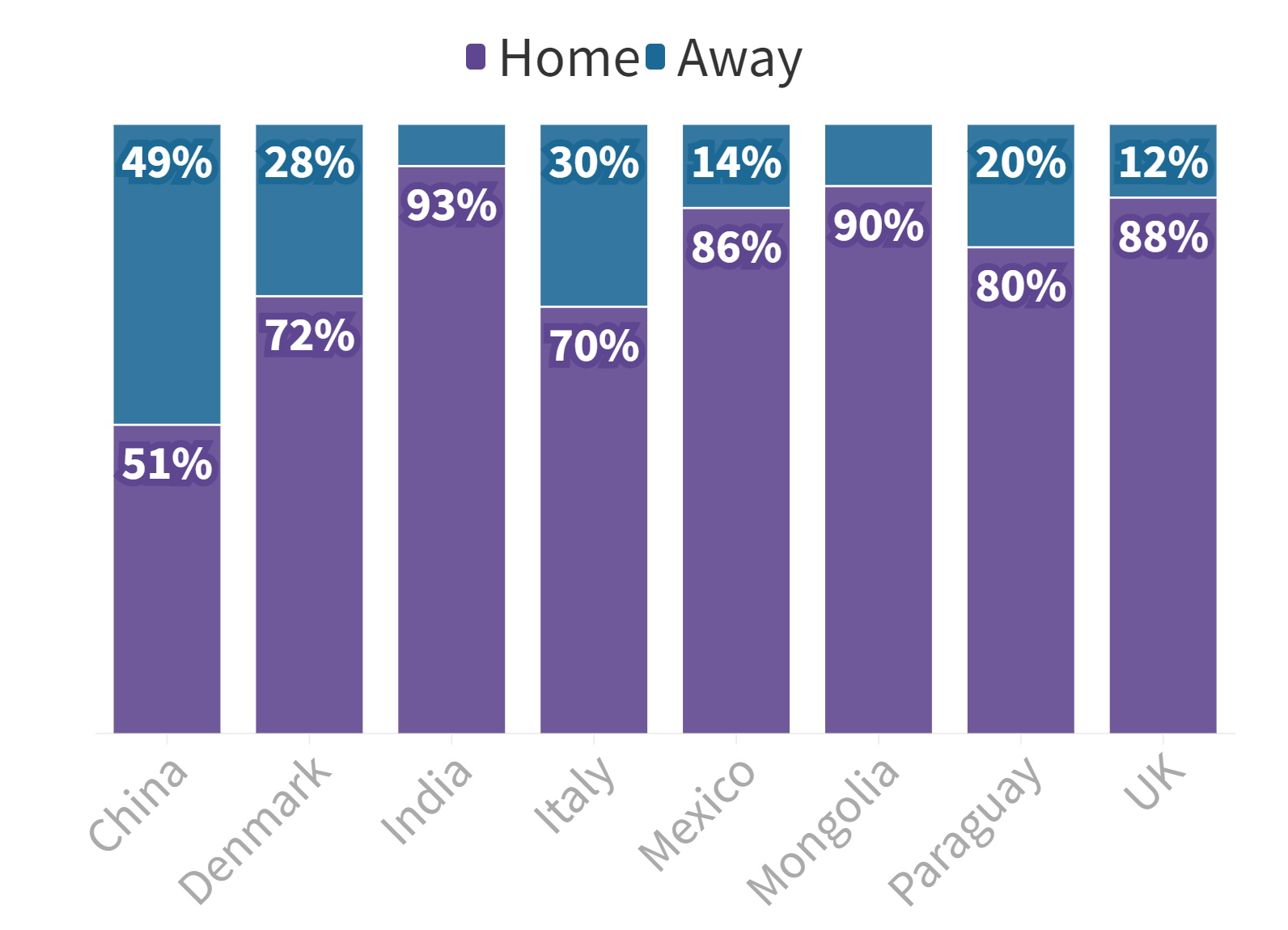}
        \caption{\centering Location context for positive mood.}
        \label{fig:pos_vs_lc}
    \end{subfigure}
    \hfill 
    \begin{subfigure}[t]{0.24\textwidth}
        \centering
        \includegraphics[width=1.125\textwidth]{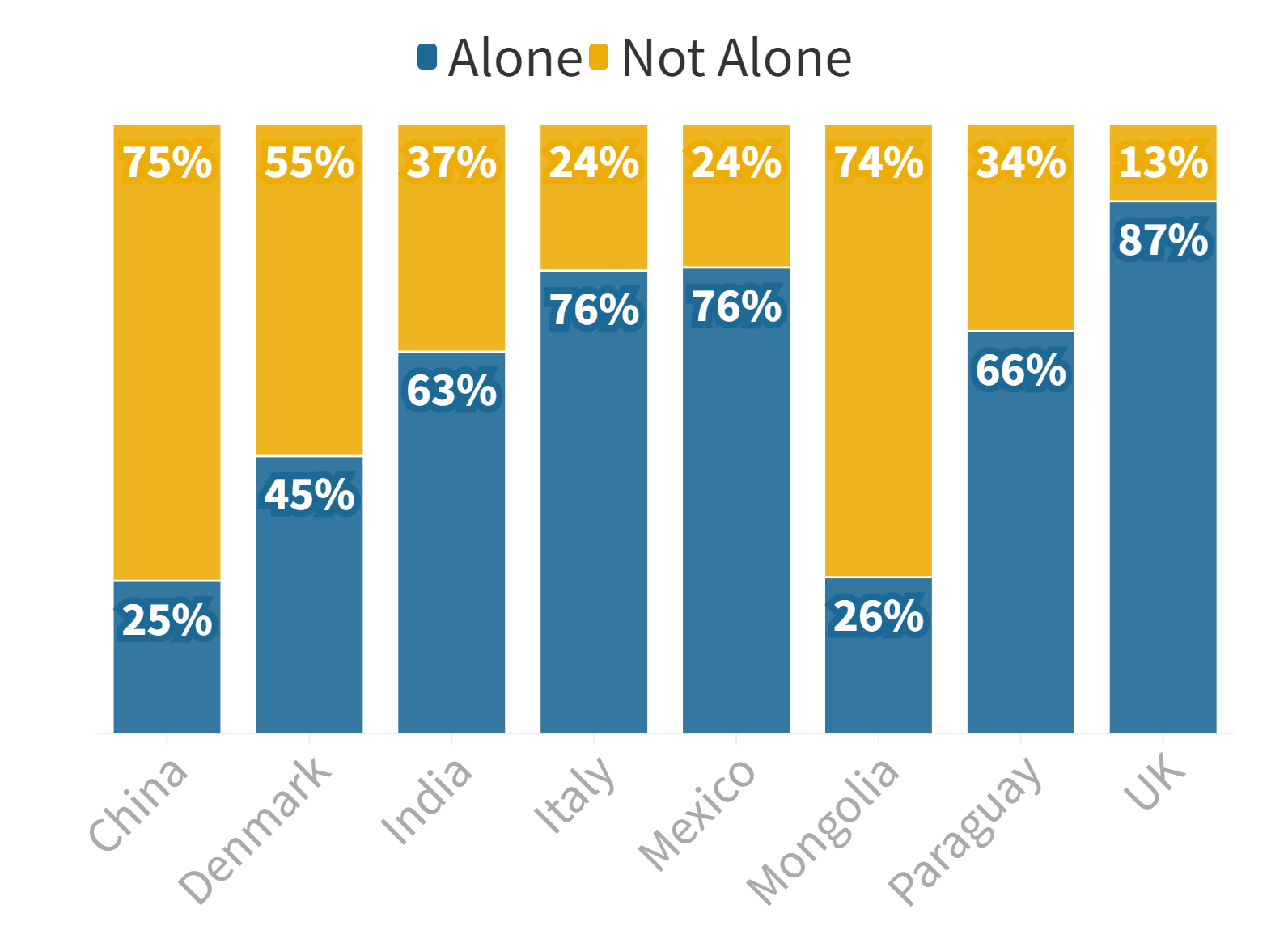}
        \caption{\centering Social context for negative mood.}
        \label{fig:neg_vs_sc}
    \end{subfigure}
    \hfill 
    \begin{subfigure}[t]{0.24\textwidth}
        \centering
        \includegraphics[width=1.125\textwidth]{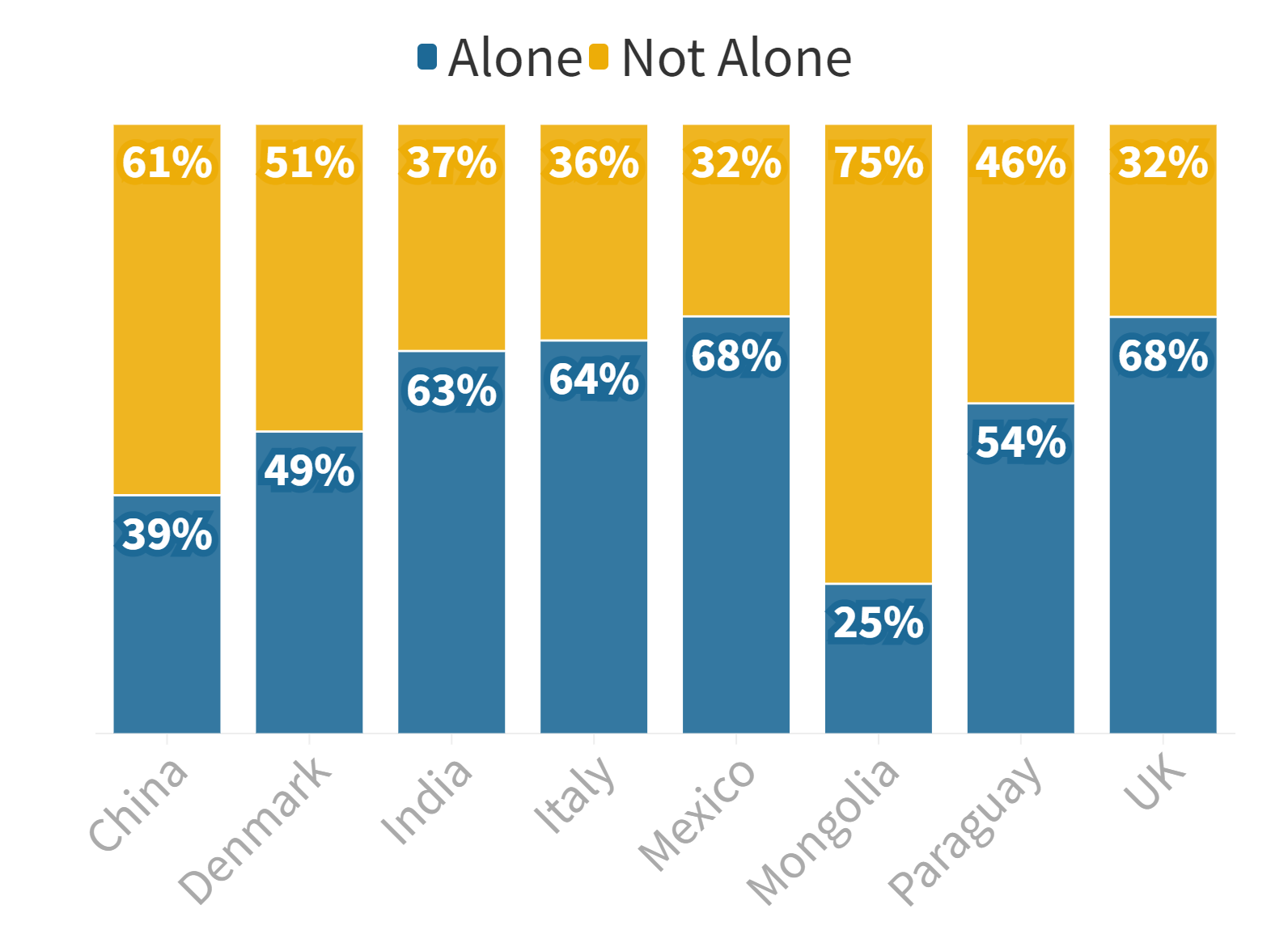}
        \caption{\centering Social context for positive mood.}
        \label{fig:pos_vs_sc}
    \end{subfigure}

    \caption{Location and social context distributions for negative and positive mood.}
    \label{fig:mood_vs_context}
\end{center}
\vspace{-0.2 in}
\end{figure*}

As mentioned in Section~\ref{subsec:mobile_app}, participants' social context and semantic location labels were captured with time diaries, in addition to mood. So, in the sub-figures of Figure~\ref{fig:mood_vs_context}, we show the distributions of social context (alone or not) and location context (home or away) for positive and negative moods. These two aspects were chosen because prior work has shown that being alone and being away from home could affect mental well-being and behavior \cite{patel2007mental, rickwood2007and, shattell2010occupational, meegahapola2021examining}. In the figure, on the X-axis, the eight countries are shown. On the Y-axis, the percentage of self-reports is shown.
Regarding location, except in China, in all other countries, most mood reports were captured when participants were home. Please note that the data was collected in the Fall of 2020, during the covid pandemic--so participants spent a significant amount of time at home. The more interesting aspect is the difference in the percentages for Positive and Negative moods: that is when comparing Figure~\ref{fig:neg_vs_lc} and Figure~\ref{fig:pos_vs_lc}. The highest difference was in Mongolia, where 67\% of negative moods were reported at home out of all negative reports. In contrast, 90\% of positive moods were reported when at home, out of all positive reports. This means that in Mongolia, participants reported a higher proportion of negative reports when away from home. This is a difference of 23\%. The difference is the lowest in Mexico. For social context, the highest difference was found in the UK, where 87\% of negative reports were done when alone. In contrast, only 68\% of positive reports were done when alone, indicating that in the UK, people tend to report more negatively when alone. The trend is similar in all other countries except China and Denmark, where proportionally more people reported that they are alone when having positive moods.

\begin{table}
    \caption{t-statistic (TS) (p-value > 0.05 : *) and Cohen's-d (CD) (all features reported here had 95\% confidence intervals not overlapping with zero) for positive, neutral, and negative moods for each country.}
    \label{tab:tstatistics}
    \resizebox{\textwidth}{!}{%
    \begin{tabular}{l l r r l r r l r r}
     
     &
     \multicolumn{3}{c}{\cellcolor[HTML]{EDEDED}\textbf{Positive}} &
     \multicolumn{3}{c}{\cellcolor[HTML]{EDEDED}\textbf{Neutral}} &
     \multicolumn{3}{c}{\cellcolor[HTML]{EDEDED}\textbf{Negative}} 
     \\

    & 
    &
    \textit{TS} &
    \textit{CD} &
    &
    \textit{TS} &
    \textit{CD} &
    &
    \textit{TS} &
    \textit{CD} 
    \\

    %
    %
    
    \multirow{5}{1.1cm}{China} &
    location altitude min & 
    10.43 & 
    0.16 & 
    
    proximity std & 
    11.51 & 
    0.17 & 
    
    app health \& fitness & 
    5.03 & 
    0.08 
    \\
    
    &
    wifi connected &
    7.98 & 
    0.11 & 
    
    location speed mean&
    7.81 & 
    0.11 & 
    
    app music \& audio &
    4.52 & 
    0.13 
    \\
    
    &
    app communication &
    7.95 & 
    0.11 & 
    
    app health \& fitness &
    7.61 & 
    0.09 & 
    
    location speed min &
    3.71 & 
    0.15 
    \\
    
    &
    screen \# of episodes &
    5.56 & 
    0.08 & 
    
    app tools &
    7.31 & 
    0.10 & 
    
    location speed mean &
    3.34 & 
    0.15 
    \\
    
    &
    app lifestyle &
    5.15 & 
    0.08 & 
    
    app personalization &
    6.87 & 
    0.10 & 
    
    proximity mean &
    2.81 & 
    0.12 
    \\

    \arrayrulecolor{Gray}
    \cmidrule{2-10}

    %
    %
    
    \multirow{5}{1.1cm}{Denmark} &
    app not found & 
    24.17 & 
    0.62 & 
    
    touch \# of events & 
    28.91 & 
    0.56 & 
    
    app puzzle & 
    8.76 & 
    0.19 
    \\
    
    &
    activity onbicycle &
    12.88 & 
    0.35 & 
    
    app video players/editors &
    28.91 & 
    0.11 & 
    
    app music \& audio &
    7.32 & 
    0.29 
    \\
    
    &
    wifi \# of devices &
    9.93 & 
    0.24 & 
    
    app weather &
    9.73 & 
    0.16 & 
    
    screen std episode &
    4.07 & 
    0.25 
    \\
    
    &
    proximity mean &
    9.89 & 
    0.27 & 
    
    app personalization &
    9.35 & 
    0.23 & 
    
    app lifestyle &
    3.84 & 
    0.11 
    \\
    
    &
    wifi std rssi &
    9.72 & 
    0.24 & 
    
    activity still &
    8.83 & 
    0.22 & 
    
    app social &
    3.21 & 
    0.18 
    \\

    \cmidrule{2-10}

    %
    %
    
    \multirow{5}{1.1cm}{India} &
    noti posted w/o duplicates & 
    6.65 & 
    0.34 & 
    
    wifi min rssi & 
    8.12 & 
    0.44 & 
    
    app business & 
    4.17 & 
    0.22 
    \\
    
    &
    wifi \# of devices &
    6.59 & 
    0.34 & 
    
    wifi mean rssi &
    6.31 & 
    0.38 & 
    
    activity tilting &
    4.02 & 
    0.28 
    \\
    
    &
    noti removed w/o duplicates &
    5.99 & 
    0.28 & 
    
    location radius of gyration &
    5.08 & 
    0.27 & 
    
    app tools &
    3.47 & 
    0.27 
    \\
    
    &
    app strategy &
    5.72 & 
    0.36 & 
    
    app books and reference &
    5.02 & 
    0.11 & 
    
    wifi min rssi &
    3.17* & 
    0.26 
    \\
    
    &
    screen \# of episodes &
    9.10 & 
    0.27 & 
    
    screen min episode &
    4.65 & 
    0.24 & 
    
    app communication &
    3.12* & 
    0.27 
    \\

    \cmidrule{2-10}

    %
    %
    
    \multirow{5}{1.1cm}{Italy} &
    proximity max & 
    26.20 & 
    0.16 & 
    
    wifi \# num of devices & 
    12.96 & 
    0.07 & 
    
    app news \& magazine & 
    11.47 & 
    0.11 
    \\
    
    &
    proximity std &
    15.19 & 
    0.09 & 
    
    app video players/editors &
    10.45 & 
    0.06 & 
    
    app action &
    8.55 & 
    0.09 
    \\
    
    &
    location speed min &
    13.21 & 
    0.08 & 
    
    activity still &
    6.80 & 
    0.04 & 
    
    activity still &
    4.98 & 
    0.07 
    \\
    
    &
    proximity mean &
    12.61 & 
    0.08 & 
    
    app adventure &
    6.52 & 
    0.03 & 
    
    app video players/editors &
    4.62 & 
    0.07 
    \\
    
    &
    wifi min rssi &
    11.75 & 
    0.07 & 
    
    app lifestyle &
    6.16 & 
    0.03 & 
    
    app social &
    4.17 & 
    0.06 
    \\

    \cmidrule{2-10}

    %
    %
    
    \multirow{5}{1.1cm}{Mexico} &
    wifi max rssi & 
    24.57 & 
    0.68 & 
    
    proximity std & 
    41.39 & 
    0.98 & 
    
    cellular lte min & 
    10.29 & 
    0.65 
    \\
    
    &
    wifi mean rssi &
    23.99 & 
    0.69 & 
    
    proximity max &
    0.93 & 
    0.06 & 
    
    wifi \# of devices &
    9.74 & 
    0.65 
    \\
    
    &
    wifi std rssi &
    22.28 & 
    0.63 & 
    
    app communication &
    20.98 & 
    0.49 & 
    
    proximity max &
    9.34 & 
    0.73 
    \\
    
    &
    screen \# of episodes &
    13.74 & 
    0.35 & 
    
    cellular lte std &
    18.23 & 
    0.32 & 
    
    app tools &
    8.79 & 
    0.73 
    \\
    
    &
    location altitude max &
    12.39 & 
    0.34 & 
    
    app music \& audio &
    18.03 & 
    0.36 & 
    
    activity still &
    7.92 & 
    0.56 
    \\

    \cmidrule{2-10}

    %
    %
    
    \multirow{5}{1.1cm}{Mongolia} &
    app not found & 
    13.76 & 
    0.12 & 
    
    wifi \# of devices & 
    16.46 & 
    0.14 & 
    
    app personalization & 
    10.99 & 
    0.19 
    \\
    
    &
    wifi std rssi &
    13.04 & 
    0.12 & 
    
    location altitude max &
    16.25 & 
    0.15 & 
    
    app music \& audio &
    10.09 & 
    0.11 
    \\
    
    &
    proximity &
    7.25 & 
    0.06 & 
    
    app role playing &
    15.24 & 
    0.09 & 
    
    app educational &
    9.79 & 
    0.05 
    \\
    
    &
    wifi connected &
    6.66 & 
    0.05 & 
    
    wifi min rssi &
    12.26 & 
    0.11 & 
    
    app sports &
    8.10 & 
    0.08 
    \\
    
    &
    user presence time &
    6.07 & 
    0.05 & 
    
    app tools &
    11.98 & 
    0.11 & 
    
    app communication &
    6.67 & 
    0.11 
    \\

    \cmidrule{2-10}

    %
    %
    
    \multirow{5}{1.1cm}{Paraguay} &
    wifi min rssi & 
    24.32 & 
    0.53 & 
    
    touch \# of events & 
    24.29 & 
    0.49 & 
    
    app role playing & 
    6.99 & 
    0.15 
    \\
    
    &
    wifi mean rssi &
    20.07 & 
    0.43 & 
    
    location speed min &
    22.49 & 
    0.48 & 
    
    app productivity &
    5.71 & 
    0.17 
    \\
    
    &
    noti posted w/o duplicates &
    13.60 & 
    0.31 & 
    
    app tools &
    12.30 & 
    0.27 & 
    
    app tools &
    5.02 & 
    0.26 
    \\
    
    &
    activity running &
    8.08 & 
    0.19 & 
    
    wifi \# of devices &
    11.53 & 
    0.25 & 
    
    screen std episode &
    4.71 & 
    0.23 
    \\
    
    &
    user presence time &
    7.69 & 
    0.17 & 
    
    app strategy &
    8.72 & 
    0.16 & 
    
    touch \# of events &
    4.49 & 
    0.22 
    \\

    \cmidrule{2-10}

    %
    %
    
    \multirow{5}{1.1cm}{UK} &
    proximity std & 
    10.52 & 
    0.18 & 
    
    wifi mean rssi & 
    17.93 & 
    0.24 & 
    
    app role playing & 
    39.48 & 
    0.53 
    \\
    
    &
    wifi \# of devices &
    10.51 & 
    0.15 & 
    
    wifi min rssi &
    15.67 & 
    0.21 & 
    
    app board &
    21.19 & 
    0.29 
    \\
    
    &
    app business &
    9.83 & 
    0.16 & 
    
    wifi max rssi &
    11.89 & 
    0.17 & 
    
    app personalization &
    17.32 & 
    0.47 
    \\
    
    &
    app tools &
    9.82 & 
    0.14 & 
    
    app role playing &
    10.34 & 
    0.13 & 
    
    touch \# of events &
    8.27 & 
    0.21 
    \\
    
    &
    proximity max &
    9.53 & 
    0.16 & 
    
    cellular lte mean &
    9.08 & 
    0.13 & 
    
    screen \# of episodes &
    6.78 & 
    0.20 
    \\
    
    \arrayrulecolor{Gray2}
    \hline 
    
    \end{tabular}%
    }
\end{table}

\begin{figure*}[t]
\begin{center}
    \begin{subfigure}[t]{\textwidth}
        \centering
        \includegraphics[width=0.7\textwidth]{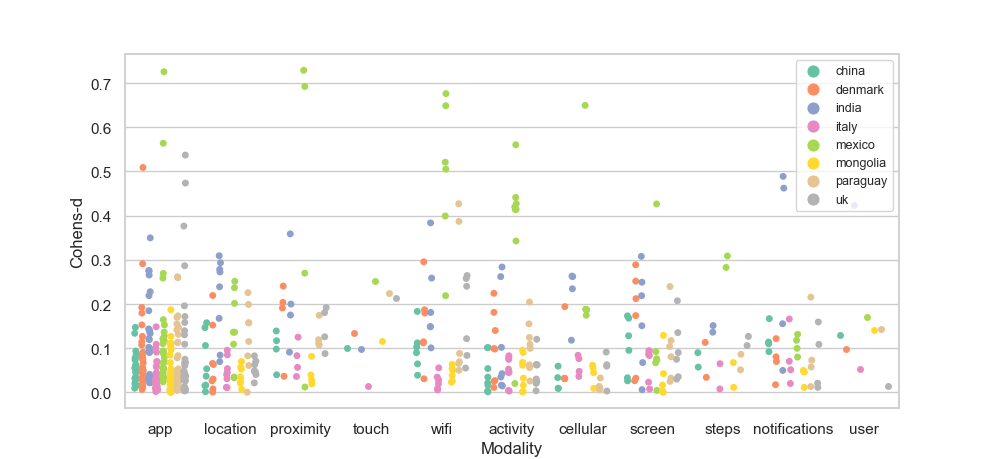}
    \end{subfigure}
    
    \caption{Cohen's-d (effect Size) distribution of features for negative and positive classes, grouped by countries and modalities.}
    \label{fig:statistics}
\end{center}
\vspace{-0.2 in}
\end{figure*}

\subsection{{Statistical Analysis.}}\label{sec:statistical_analysis}

In this section, we seek to understand features with high statistical significance in discriminating either positive, neutral, or negative classes from the other two. Therefore, in Table~\ref{tab:tstatistics}, we show the t-statistic \cite{Kim2015} and p-values \cite{Greenland2016} (p-values higher than 0.05 after Bonferroni correction for multiple hypothesis testing \cite{weisstein2004bonferroni} are marked with *). In addition, since p-values are limited in determining statistical significance \cite{Lee2016}, we also report Cohen's-d \cite{Rice2005} (all features have 95\% confidence interval not crossing zero \cite{Lakens2013}) for positive, neutral, and negative classes for each country. The rule of thumb to evaluate Cohen's-d is 0.2 = small effect size, 0.5 = medium effect size, and 0.8 = large effect size. For the positive mood, across all the European countries and Mongolia, proximity sensor-related features were among the top five features, indicating that phone usage and/or location of the phone could reveal positive moods. However, except for Denmark, where there was a small effect size, the proximity feature had less than small effect sizes in all other countries. In addition, in Denmark, cycling activity was indicative of positive moods. Interestingly, Copenhagen in Denmark is a city widely known for cycling \cite{pucher2007frontiers}, which might explain this finding. Further, running activity could discriminate positive mood with a small effect size in Paraguay. Prior work has also shown that high physical activity could lead to positive moods and less stress \cite{biddle2000emotion, kanning2010active}.  

Regarding the negative class, app features were predominant in most countries. For some apps, high usage indicated negative moods (e.g., puzzle in Denmark, news \& magazine in Italy, etc.). In contrast, for some apps, low usage indicated negative moods (e.g., health \& fitness in China, music \& audio in China and Mongolia, role-playing games in UK and Paraguay, etc.). In addition, for both UK and Paraguay, a high number of touch events on the phone was indicative of negative moods. This finding is generally in line with prior studies that examined fine-grained smartphone usage and mental well-being \cite{hung2016predicting, likamwa2013moodscope}. In summary, features from modalities such as app usage, screen and phone usage events (episodes, touch events, user presence, proximity, etc.), WiFi, activity types, and location were among the ones that helped discriminate between different moods. Further, except for the `proximity std' feature in Mexico for neutral mood, none of the features had a larger effect size. For a few country-mood pairs, there were cases of features having above medium effect sizes (e.g., number of touch events in Denmark for Neutral, many features from modalities such as cellular, WiFi, proximity, etc. in Mexico, minimum RSSI value for WiFi in Paraguay for Positive, and role-playing apps in the UK for Negative). Figure~\ref{fig:statistics} shows the distribution of Cohen's-d values for all features grouped by sensing modalities for the two classes studied in this paper (i.e., negative vs. positive). Results indicate that depending on the country, the expressiveness of different sensing modalities in discriminating negative classes from other classes is different. For example, for `app' features, effect sizes are small for countries such as China, Italy, and Mongolia. In contrast, more informative features with larger effect sizes are present for Denmark, Mexico, and the UK.

\section{Mood Inference (RQ2 \& RQ3)}\label{sec:inference}

\subsection{Experimental Setup}

The primary goal of this paper is to investigate aspects related to mood inference, personalization, and generalization to different countries using smartphone sensing data. As described and defined in Figure~\ref{fig:overview} and Table~\ref{tab:terminology}, we use two model types: population-level and hybrid, to examine personalization to individuals, and four modeling approaches: country-specific, continent-specific, country-agnostic, and multi-country, to examine generalization and country-wise performance. Hence, this section will describe the operationalization of the experimental protocol. 

We used python with scikit-learn \cite{pedregosa2011scikit} and Keras \cite{chollet2015keras} frameworks to conduct all experiments. Initially, we conducted country-specific experiments with different model types such as random forest (RF), gradient boosting, support vector classification, XGBoost, AdaBoost, and multi-layer perceptron neural networks \cite{rf_inbook, rish2001empirical, natekin2013gradient, chen2016xgboost, schapire2013explaining, noble2006support}. We obtained the best results for a larger majority of inferences with RFs. In addition, these models allow interpreting results better because they provide Gini feature importance values for trained models. Because of these reasons and space limitations, we will only report results for RF models with default parameters in this paper \footnote{Note that we also tried out GridSearch for parameters in the random forest (for n\_estimators: 50, 100-2000 with intervals of 100, max\_depth: 2-16 with intervals of 2, min\_samples\_split: 2-10) that did not yield better performance than the default parameters (n\_estimators: 100, max\_depth: NA, min\_samples\_split: 2), except in a few cases. Hence, we used default parameters for all experiments for consistency.}. Further, to fill in missing values of the dataset, we used k-nearest-neighbor (kNN) imputation \cite{beretta2016nearest, zhou2018missing}. In addition, we report all the results with the area under the receiver operating characteristic curve (AUROC) \cite{bradley1997use} because they provide a better assessment of performance when dealing with imbalanced data (when used with macro averaging which gives equal emphasis to all classes in an inference). While we provided a basic description of model types in Table~\ref{tab:terminology}, the operationalization of models is given below. 

\begin{itemize}
    \item \textbf{Population-Level Models (PLM)}: Since this represents a scenario where models are deployed to a set of users unseen in model training, we use the leave-n-participants-out strategy when testing models. This is an extension of leave-one-out cross-validation, where we consider $n$ users in testing instead of one. Hence, if the number of users in the considered population is $N$, we pick $n$ such that it is roughly 20\% of $N$ (can be obtained with group-k-fold cross-validation with k = 5 in scikit-learn). So, for each $n$ user in the testing split, 50\% of their data would be used for testing to be coherent with hybrid models (stratified based on users and mood labels), and data from the rest of the $N-n$ users would be used for the training split. Then, experiments were repeated ten times by randomly sampling $n$ users, and the results were averaged. 
    
    \item \textbf{Hybrid Models (HM)}: Since this represents a scenario where models are deployed to a set of users already seen in model training (hence partially personalized models), we first use the leave-n-participants-out strategy similar to PLM. So, for each $n$ user in the testing split, data from the rest of the $N-n$ users would be used for the training split. In addition, 50\% of the data from the testing split (stratified based on users and mood labels) would be included in the training set to represent partial personalization. In addition, an equal number of data points to the number of data points added to the training set from the testing set would be removed randomly to make the number of data points in the training and testing sets for HM and PLM equal making them more comparable. Finally, experiments were repeated ten times by randomly sampling $n$ users, and the results were averaged. 
\end{itemize}

Using the above two model types, we conducted the experiments using four approaches. The country-specific approach examines how models trained within a country perform. We examine both PLM and HM types for this approach, hence examining the personalization within countries. The country-agnostic approach examines how models trained in one or a few countries generalize to a new country. With PLM and HM model types, we examine how personalization affects model performance when models are deployed to countries unseen on training data. The multi-country approach is similar to a one-size-fits-all model trained with data from all available countries. This is similar to a model in which country diversity is ignored. Both PLM and HM model types were used to examine the effects of personalization on model performance.

\begin{table}
    \caption{{Country-Specific and Multi-Country results with PLM and HM: }Mean ($\bar{S}$) and Standard Deviation ($S_\sigma$) AUROC scores computed from ten iterations. Results are presented as $\bar{S} (S_\sigma)$, where $S$ is AUROC.}
    \label{tab:specific-results}
    \resizebox{0.65\textwidth}{!}{%
    \begin{tabular}{l l l l l}

    &
    \multicolumn{2}{c}{\cellcolor[HTML]{EDEDED}\textbf{PLM}}&
    \multicolumn{2}{c}{\cellcolor[HTML]{EDEDED}\textbf{HM}} \\
    
    &
    \multicolumn{1}{c}{\cellcolor[HTML]{D9D9D9}\textbf{Two-Class}} &
    \multicolumn{1}{c}{\cellcolor[HTML]{D9D9D9}\textbf{Three-Class}} &
    \multicolumn{1}{c}{\cellcolor[HTML]{D9D9D9}\textbf{Two-Class}} &
    \multicolumn{1}{c}{\cellcolor[HTML]{D9D9D9}\textbf{Three-Class}} 
    \\
    
    \arrayrulecolor{Gray}
    \hline
    
    
    Baseline &
    .50 (.00) & 
    .50 (.00) & 
    .50 (.00) & 
    .50 (.00)  
    \\
    
    \hline 
    
    China &
    .51 (.04) & 
    .45 (.04) & 
    .78 (.02) & 
    .79 (.01)  
    \\
    
    Denmark &
    .41 (.10) & 
    .56 (.03) & 
    .83 (.03) & 
    .86 (.01)  
    \\
    
    India &
    .46 (.15) & 
    .45 (.04) & 
    .79 (.03) & 
    .76 (.02)  
    \\
    
    Italy &
    .55 (.05) & 
    .52 (.01) & 
    .82 (.01) & 
    .81 (.00)  
    \\
    
    Mexico &
    .62 (.21) & 
    .62 (.13) & 
    .98 (.01) & 
    .94 (.01) 
    
    \\
    
    Mongolia &
    .49 (.08) & 
    .49 (.02) & 
    .85 (.01) & 
    .83 (.00) 
    \\
    
    Paraguay &
    .48 (.08) & 
    .53 (.01) & 
    .84 (.01) & 
    .84 (.01) 
    \\
    
    UK &
    .56 (.05) & 
    .52 (.05) & 
    .91 (.01) & 
    .87 (.00) 
    \\
    
    \hline 
    
    Aggregate &
    .51 (.10) & 
    .52 (.04) & 
    .85 (.02) & 
    .84 (.01) 
    \\

    
    Multi-Country &
    .52 (.03) & 
    .53 (.02) & 
    .83 (.01) & 
    .79 (.00) 
    \\
    
    Multi-Country (Balanced) &
    .53 (.02) & 
    .52 (.03) & 
    .81 (.03) & 
    .78 (.02) 
    \\
    
    \arrayrulecolor{Gray2}
    \hline

    \end{tabular}%
    }
\end{table}

\subsection{Results}

\subsubsection{Country-Specific Models} In Table~\ref{tab:specific-results}, we show country-specific results with PLM and HM. In addition, we also show the aggregate results from country-specific (as `Aggregate') and multi-country models. Under `Multi-Country (Balanced)', we use an equal number of data points from each country (equal to the country with the minimum number of data points, which is India) by randomly sampling when training and testing models. The results show that PLMs do not perform well for two and three-class inferences. Models in Mexico performed better than in other countries. These results are reasonable because many features in Mexico had medium to large effect sizes, as shown in Figure~\ref{fig:statistics}. However, HM results show that they perform better than PLMs, showing the usefulness of personalization within each country. With HMs, the performance for two-class inference almost doubled for Denmark, and even for other countries, the AUROC bump was above 30\%. These results suggest that for both two-class and three-class inferences, partial personalization within each country leads to significant improvements in performance. When the aggregate results of country-specific models are compared with multi-country models, PLMs do not show a significant difference. However, with HMs, it is clear that country-specific models outperform multi-country models by 2\% for two-class and 5\% for three-class. This suggests that model personalization within countries leads to better performance when compared to the personalization of one-size-fits-all models. This is reasonable given that we are reducing the distributional shift by only considering data within a country and adding an effect of personalization by being geographically diversity-aware. In addition, the `Multi-Country' approach performed slightly better than the `Multi-Country (Balanced)' case. This could be because, in the imbalanced case, models favor countries with more data points, such as Italy and Mongolia, leading to a slight increase in performance for those countries that occupy a majority of the dataset. Furthermore, regardless of whether it is a two/three-class inference, the performance of models did not degrade much.

\subsubsection{Country-Agnostic I Models} 

Next, we examine the country-agnostic approach. Table~\ref{tab:agnostic-two-class} and Table~\ref{tab:agnostic-three-class} show the results for two-class and three-class inferences, respectively. In both tables, we first show results for models trained in specific countries when tested on an unseen country in the form of a matrix with an empty diagonal. Then, under `Aggregate', we show the aggregate value of those results for each training country (e.g., PLM performance for models trained in China when deployed to other countries). In addition, we calculated AUROC scores for the same set of models with partial personalization (all the results are not shown here due to space limitations), and similar to the aggregate of PM, we show the aggregate values under HM. Results show that PLMs do not generalize well to new countries with AUROCs of 0.47 - 0.52. However, these results are on par with PLM accuracies in country-specific and multi-country approaches. This suggests that regardless of the country from where sensing data were obtained to train models for mood inference, PLMs performed similarly. However, HM results convey an opposite conclusion for two and three-class inferences. For the two-class inference, the country-specific approach had AUROC scores in the range of 0.78-0.98, whereas the country-agnostic approach yielded scores in the range of 0.66-0.73. A similar pattern can be seen for three-class inference, where scores dropped from 0.76-0.94 to 0.65-0.71. This shows that the effect of personalization achieved with HMs is strong for the country-specific approach, whereas country-agnostic models still did not generalize well. However, we also noticed that with HMs for both two-class and three-class inferences, models trained in European countries consistently performed better in other European countries than the rest. For example, in the two-class inference, the Italian model had AUROC scores of 0.76 and 0.78 in Denmark and the UK, respectively. In contrast, the next best score for the Italian model was 0.70 in India. Finally, for three-class inference, the UK model had AUROC scores of 0.73 and 0.75 for Italy and Denmark, respectively, whereas the next best score was 0.69 for Paraguay. These results could be partly justified given that European countries have somewhat closer everyday patterns that could get captured in the models.

\begin{table}
    \caption{{Country-Agnostic I PLM \& HM: }Two-Class Inference -- Mean ($\bar{S}$) and Standard Deviation ($S_\sigma$) of AUROC scores obtained by testing each Country-Specific model (rows) on a new country. Results are presented as $\bar{S} (S_\sigma)$, where $S$ is AUROC score. Aggregate of the reported population-level results and results from hybrid models indicated under `Aggregate'.}
    \label{tab:agnostic-two-class}
    \resizebox{0.95\textwidth}{!}{%
    \begin{tabular}{lrrrrrrrr|rr}
    
    &
    \multicolumn{8}{c}{\cellcolor[HTML]{EDEDED}\textbf{Testing (PLM)}} &
    \multicolumn{2}{c}{\cellcolor[HTML]{EDEDED}\textbf{Aggregate}} 
    \\
     
    \rowcolor[HTML]{D9D9D9} 
    \cellcolor[HTML]{EDEDED} \textbf{Training}  &
    \multicolumn{1}{c}{\cellcolor[HTML]{D9D9D9}\textbf{China}} & \multicolumn{1}{c}{\cellcolor[HTML]{D9D9D9}\textbf{Denmark}} & \multicolumn{1}{c}{\cellcolor[HTML]{D9D9D9}\textbf{India}} & \multicolumn{1}{c}{\cellcolor[HTML]{D9D9D9}\textbf{Italy}} &
    \multicolumn{1}{c}{\cellcolor[HTML]{D9D9D9}\textbf{Mexico}} &
    \multicolumn{1}{c}{\cellcolor[HTML]{D9D9D9}\textbf{Mongolia}} & \multicolumn{1}{c}{\cellcolor[HTML]{D9D9D9}\textbf{Paraguay}} & \multicolumn{1}{c}{\cellcolor[HTML]{D9D9D9}\textbf{UK}} & \multicolumn{1}{c}{\cellcolor[HTML]{D9D9D9}\textbf{PLM}} & \multicolumn{1}{c}{\cellcolor[HTML]{D9D9D9}\textbf{HM}}
    \\
     
    \arrayrulecolor{Gray}

    \hline
     
    China &
      &
    .53 (.02) &
    .44 (.03) &
    .49 (.01) &
    .58 (.05) &
    .50 (.01) &
    .42 (.03) &
    .51 (.02) &
    .55 (.02) &
    .67 (.04)
    \\
    
    Denmark &
    .51 (.00) &
    &
    .47 (.01) &
    .51 (.00) &
    .58 (.02) &
    .50 (.00) &
    .58 (.01) &
    .46 (.00) &
    .52 (.01) &
    .69 (.03)
    \\
    
    India &
    .48 (.00) &
    .37 (.00) &
     &
    .50 (.00) &
    .40 (.02) &
    .50 (.00) &
    .44 (.01) &
    .52 (.00) &
    .46 (.00) &
    .70 (.02) 
    \\

    Italy &
    .49 (.00) &
    .45 (.00) &
    .51 (.01) &
    &
    .40 (.02) &
    .51 (.01) & 
    .48 (.00) &
    .50 (.00) &
    .48 (.01) &
    .69 (.02) 
    \\
    
    Mexico &
    .49 (.00) &
    .58 (.01) &
    .44 (.01) &
    .49 (.00) &
     & 
    .49 (.01) &
    .56 (.01) &
    .47 (.01) &
    .50 (.01) &
    .73 (.03) 
    \\
    
    Mongolia &
    .49 (.00) &
    .48 (.01) &
    .52 (.00) &
    .50 (.00) &
    .51 (.00) &
     &
    .48 (.00) &
    .51 (.00) &
    .50 (.00) &
    .71 (.03) 
    \\
    
    Paraguay &
    .51 (.00) &
    .53 (.01) & 
    .49 (.01) &
    .50 (.00) &
    .55 (.02) & 
    .53 (.01) &
    &
    .50 (.01) &
    .52 (.01) &
    .70 (.02) 
    \\

    UK &
    .48 (.01) &
    .43 (.02) &
    .57 (.00) &
    .50 (.01) &
    .32 (.01) &
    .50 (.01) &
    .49 (.01) &
    & 
    .47 (.01) &
    .66 (.02) 
    \\
    
    \arrayrulecolor{Gray2}
    
    \hline
    \end{tabular}%
    }
\end{table}

\begin{table}
    \caption{{Country-Agnostic I PLM \& HM: }Three-Class Inference -- Mean ($\bar{S}$) and Standard Deviation ($S_\sigma$) of AUROC scores obtained by testing each Country-Specific model (rows) on a new country. Results are presented as $\bar{S} (S_\sigma)$, where $S$ is AUROC score. Aggregate of the reported population-level results and results from hybrid models indicated under `Aggregate'.}
    \label{tab:agnostic-three-class}
    \resizebox{0.95\textwidth}{!}{%
    \begin{tabular}{lrrrrrrrr|rr}
    
     &
    \multicolumn{8}{c}{\cellcolor[HTML]{EDEDED}\textbf{Testing (PLM)}} &
    \multicolumn{2}{c}{\cellcolor[HTML]{EDEDED}\textbf{Aggregate}}  
     \\
     
     \rowcolor[HTML]{D9D9D9} 
     \cellcolor[HTML]{EDEDED} \textbf{Training}  &
     \multicolumn{1}{c}{\cellcolor[HTML]{D9D9D9}\textbf{China}} & \multicolumn{1}{c}{\cellcolor[HTML]{D9D9D9}\textbf{Denmark}} & \multicolumn{1}{c}{\cellcolor[HTML]{D9D9D9}\textbf{India}} & \multicolumn{1}{c}{\cellcolor[HTML]{D9D9D9}\textbf{Italy}}  &
     \multicolumn{1}{c}{\cellcolor[HTML]{D9D9D9}\textbf{Mexico}} &
     \multicolumn{1}{c}{\cellcolor[HTML]{D9D9D9}\textbf{Mongolia}} & \multicolumn{1}{c}{\cellcolor[HTML]{D9D9D9}\textbf{Paraguay}} & \multicolumn{1}{c}{\cellcolor[HTML]{D9D9D9}\textbf{UK}} & \multicolumn{1}{c}{\cellcolor[HTML]{D9D9D9}\textbf{PLM}} & \multicolumn{1}{c}{\cellcolor[HTML]{D9D9D9}\textbf{HM}}
     \\

     \arrayrulecolor{Gray}
    \hline

    China &
     &
    .48 (.01) &
    .54 (.01) &
    .48 (.01) &
    .47 (.01) &
    .50 (.01) &
    .51 (.01) &
    .50 (.00) &
    .50 (.01) &
    .68 (.02) 
    \\
    
    Denmark &
    .52 (.01) &
    &
    .41 (.02) &
    .56 (.01) &
    .54 (.04) &
    .51 (.01) &
    .50 (.02) &
    .58 (.01) &
    .52 (.02) &
    .66 (.04) 
    \\
    
    India &
    .52 (.01) &
    .42 (.02) &
     &
    .52 (.01) &
    .38 (.02) &
    .52 (.01) &
    .52 (.01) &
    .38 (.01) &
    .47 (.01) &
    .68 (.03) 
    \\
    
    Italy &
    .52 (.01) &
    .49 (.01) &
    .47 (.02) &
    &
    .32 (.02) &
    .51 (.01) & 
    .51 (.00) &
    .54 (.00) &
    .48 (.01) &
    .69 (.02) 
    \\
    
    Mexico &
    .49 (.00) &
    .59 (.00) &
    .44 (.00) &
    .47 (.00) &
    & 
    .50 (.00) &
    .61 (.00) &
    .54 (.00) &
    .52 (.00) &
    .71 (.02) 
    \\
    
    Mongolia &
    .49 (.00) &
    .50 (.00) &
    .43 (.00) &
    .51 (.00) &
    .55 (.00) &
    &
    .54 (.00) &
    .53 (.00) &
    .51 (.00) &
    .67 (.02) 
    \\
    
    Paraguay &
    .44 (.01) &
    .51 (.02) & 
    .48 (.03) &
    .52 (.01) &
    .58 (.05) & 
    .53 (.01) &
    &
    .55 (.01) &
    .52 (.02) &
    .65 (.04) 
    \\
    
    UK &
    .53 (.01) &
    .51 (.01) &
    .51 (.03) &
    .53 (.01) &
    .40 (.06) &
    .52 (.01) &
    .53 (.02) &
    &
    .50 (.02) &
    .67 (.03) 
    \\

    \arrayrulecolor{Gray2}
    \hline
    \end{tabular}%
    }
\end{table}

\subsubsection{{Country-Agnostic II Models}} In Table~\ref{tab:agnostic-ii}, we show results for country-agnostic models that were trained in seven countries and tested in the shown country. Compared to the previous setting, where the models were trained in only one country and tested in another, these models capture a more considerable intra-subject variability in model training. Moreover, HM results were not included here because, technically, it is similar to the HM of multi-country models. PLM results show that the performance is not high for both two-class and three-class inferences. For some countries, performance slightly increased compared to country-specific (e.g., China, Paraguay in two-class). For some, the performance declined (e.g., India, Denmark, Italy, Mexico, and the UK in two-class). Hence, there is no clear evidence that having more data from multiple countries would help to generalize better for an unseen country, even in this case.

\begin{table}
    \caption{{Country-Agnostic II PLM: }Mean ($\bar{S}$) and Standard Deviation ($S_\sigma$) of AUROC scores obtained by testing each a seven-country model on data from a new country. Results are presented as $\bar{S} (S_\sigma)$, where $S$ is the AUROC.}    
    \label{tab:agnostic-ii}
     \resizebox{0.35\textwidth}{!}{%
    \begin{tabular}{l l l}

    &
    \multicolumn{1}{c}{\cellcolor[HTML]{EDEDED}\textbf{Two-Class}} &
    \multicolumn{1}{c}{\cellcolor[HTML]{EDEDED}\textbf{Three-Class}} 
    \\
    \arrayrulecolor{Gray}
    \hline
    
    Baseline &
    .50 (.00) &
    .50 (.00) 
    \\
    
    \hline 
    
    China &
    .54 (.01) &
    .48 (.01) 
    \\
    
    Denmark & 
    .51 (.02) &
    .48 (.01) 
    \\
    
    India &
    .53 (.03) &
    .47 (.01)
    \\
    
    Italy &
    .54 (.01) &
    .50 (.01) 
    \\
    
    Mexico &
    .41 (.02) &
    .54 (.01) 
    \\ 
    
    Mongolia &
    .49 (.01) &
    .49 (.01) 
    \\
    
    Paraguay &
    .56 (.01) &
    .55 (.01) 
    \\
    
    UK &
    .48 (.01) &
    .51 (.01)  
    \\ 
    
    \arrayrulecolor{Gray2}
    \hline
    \end{tabular}
    }
\end{table}

\begin{table}
    \caption{{Multi-Country and Continent-Specific with PLM and HM: }Mean ($\bar{S}$) and Standard Deviation ($S_\sigma$) of F1-scores and AUROC scores obtained by testing the "worldwide" model. Results are presented as $\bar{S} (S_\sigma)$, where $S$ is any of the two
    metrics.}    
    \label{tab:multi-country}
    
    \resizebox{0.7\textwidth}{!}{%
    \begin{tabular}{l l l l l}

    &
    \multicolumn{2}{c}{\cellcolor[HTML]{EDEDED}\textbf{PLM}}&
    \multicolumn{2}{c}{\cellcolor[HTML]{EDEDED}\textbf{HM}} \\
    
    &
    \multicolumn{1}{c}{\cellcolor[HTML]{D9D9D9}\textbf{Two-Class}} &
    \multicolumn{1}{c}{\cellcolor[HTML]{D9D9D9}\textbf{Three-Class}} &
    \multicolumn{1}{c}{\cellcolor[HTML]{D9D9D9}\textbf{Two-Class}} &
    \multicolumn{1}{c}{\cellcolor[HTML]{D9D9D9}\textbf{Three-Class}} 
    \\
    
    \arrayrulecolor{Gray}
    \hline

    Baseline &
    .50 (.00) & 
    .50 (.00) & 
    .50 (.00) & 
    .50 (.00)  
    \\
    
    \hline 
    
    Europe &
    .58 (.03) & 
    .50 (.03) & 
    .89 (.03) & 
    .86 (.02)  
    \\
    
    Asia &
    .51 (.02) & 
    .52 (.05) & 
    .79 (.03) & 
    .74 (.01)  
    \\
    
    Multi-Country &
    .52 (.03) & 
    .53 (.02) & 
    .83 (.01) & 
    .86 (.03)  
    \\
    
    \hline 
    
    Europe (Balanced) &
    .53 (.02) & 
    .50 (.05) & 
    .86 (.04) & 
    .82 (.03)  
    \\
    
    Asia (Balanced) &
    .52 (.04) & 
    .54 (.03) & 
    .79 (.02) & 
    .76 (.02)  
    \\
    
    Multi-Country (Balanced) &
    .53 (.02) & 
    .52 (.03) & 
    .81 (.03) & 
    .78 (.02) 
    \\
    
    \arrayrulecolor{Gray2}
    \hline

    \end{tabular}%
    }
    
\end{table}

\subsubsection{Multi-Country and Continent-Specific Models} 

Finally, in Table~\ref{tab:multi-country}, we show the results for the multi-country approach and also the continent-specific approach that is similar to the country-specific; however, instead of countries, we considered two continents: Europe (Italy, Denmark, UK) and Asia (China, Mongolia, India) \footnote{There are arguments for and against on whether North and South America are a single continent or two \cite{nationsonline2022continents, worldometer20227continents, Ward2020How}. In the Anglo-Saxon world, it is often stated that there are seven continents, with North and South America being separate. In contrast, it is taught otherwise in Latin America \cite{Ward2020How}. Hence, we did not include ‘America’ results by combining Mexico and Paraguay.}. The primary motivation for examining these models is the result we obtained in the country-agnostic approach, where for HM, models trained in European countries performed better in other European countries with HMs. Results for the continent-specific approach show that models performed similarly to any other approach for both two-class and three-class inferences for PLM. However, the Europe model for two-class inference had an AUROC score of 0.58, which is second only to the Mexican model (0.62) in the country-specific approach.

Furthermore, results show that the continent-specific model for Europe with an AUROC of 0.89 for two-class inference, performed better than the multi-country (0.83) and even country-specific approach for Italy (0.82) and Denmark (0.83) and closer to the country-specific UK model with an AUROC of 0.91. Similar results can be seen for three-class HM inference. This suggests that for western Europe, where everyday patterns might be somewhat similar across countries, continent-specific models could perform reasonably. However, for the continent-specific Asian model, it is not the same. For example, for the two-class inference, the Asia model had an AUROC score of 0.79, which is similar to country-specific China (0.78) and India (0.79) results but significantly lower than the result for Mongolia (0.85).
On the other hand, for the three-class HM, the Asia approach reached an AUROC of 0.74, whereas China, India, and Mongolia models reached 0.79, 0.76, and 0.84, respectively. Hence, continent-specific models did not perform as well as country-specific or multi-country models for Asia. This could be because even though China, India, and Mongolia are geographically on the same continent, the behaviors and cultures of students are different. In addition, ‘balanced’ models decreased performance for Europe and Multi-Country, whereas for Asia, it is not the same, where three-class HM performance increased in the balanced case. Again, this is because India and China get more representation in training, leading to better performance in testing.

\begin{figure}
    \includegraphics[width=\textwidth]{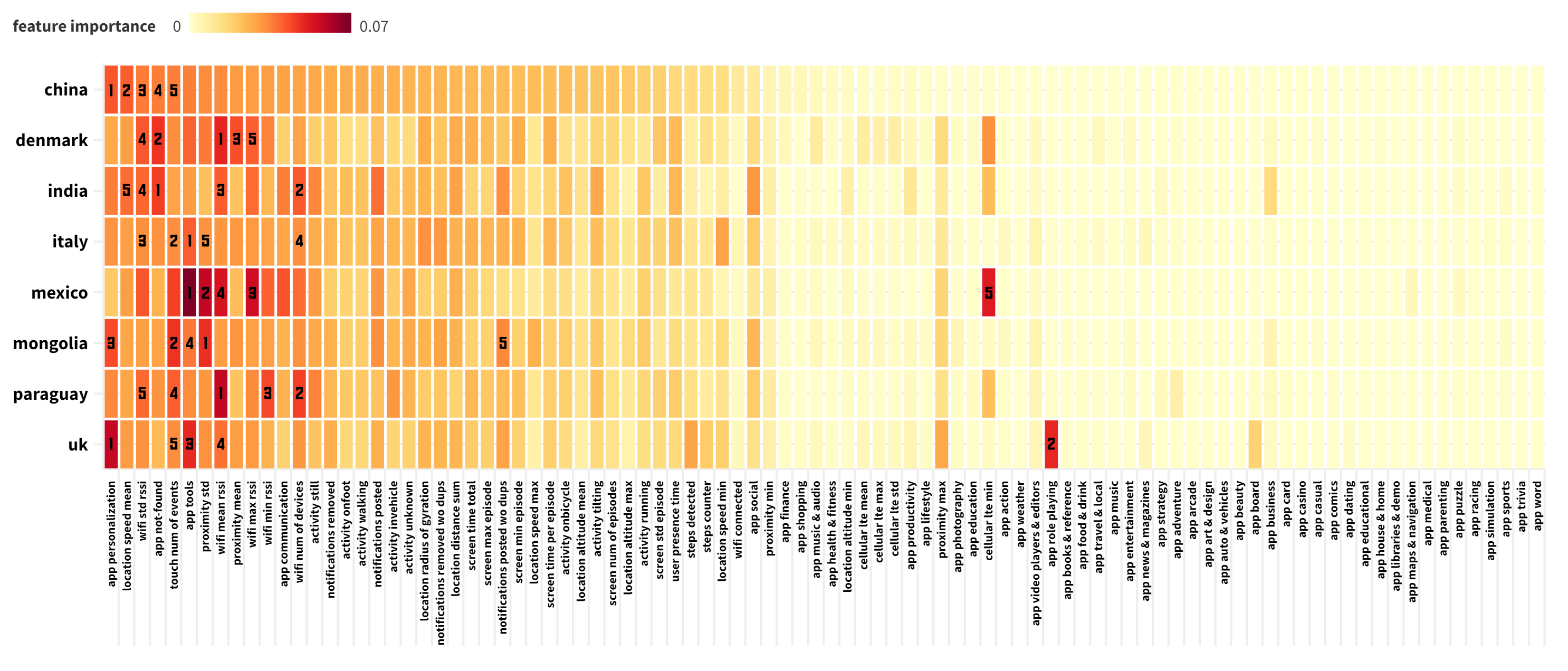}
    \caption{{Country-Specific HM: } Gini feature importance values from RF models for two-class inference.}
    \label{fig:fi_two_class}
\end{figure}

\begin{figure}
    \includegraphics[width=\textwidth]{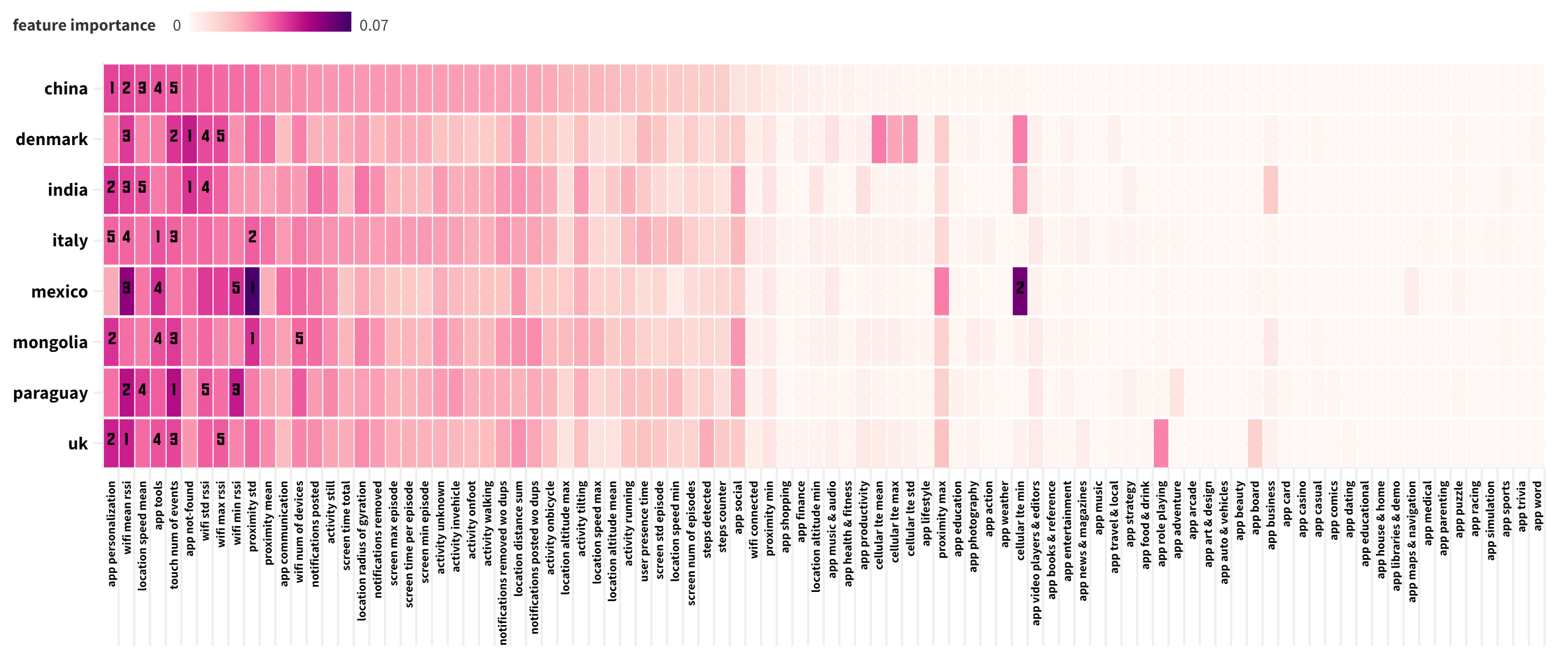}
    \caption{{Country-Specific HM: }Gini feature importance values from RF models for three-class inference.}
    \label{fig:fi_three_class}
\end{figure}

\subsubsection{Gini Feature Importance Values}

Figure~\ref{fig:fi_two_class} and Figure~\ref{fig:fi_three_class} show the Gini feature importance values for each country for two-class and three-class mood inferences with HMs. We report diagrams for HMs because they provide the highest performance. Further, the top five features within each country are marked with numbers from one to five. Moreover, in both diagrams, values are arranged in the decreasing order of values in China, from left to right. For both inferences, many apps had very low feature importance values.
On the other hand, `app personalization' and `app tools' were among the top five features for many countries. For the UK, personalization apps were highly important in two and three-class inferences. However, for Mexico, the importance of the feature was relatively lower in both inferences. In addition, the number of touch events on the phone was within the top five features for Italy, Mongolia, Paraguay, and the UK in the two-class inference and all countries except India and Mexico in the three-class inference. This aligns with previous literature that presented findings of typing and touch events indicative of aspects such as mood and stress \cite{likamwa2013moodscope}. Another feature discussed in the literature on psychological aspects and mobile sensing \cite{canzian2015trajectories}, which appeared again in the diagrams is speed, calculated using location sensors (`location speed mean'). Diagrams indicate that the feature was in the top five in two-class inference for India and China and three-class inference for India and Paraguay. In addition to these features, multiple features captured using Wifi signals were among the top five in all countries. Wifi-related features (i.e., `wifi std rssi',`wifi mean rssi', `wifi min rssi', `wifi max rssi' - The standard deviation/mean/minimum/maximum of RSSI signal strengths captured with unique devices within the time window) were present with high importance values for all countries across both inferences. Prior work highlights that the number of wifi devices and signal strengths could be indicative of user context, including the location \cite{santani2018drinksense}, and location-related features have shown to be closely tied to the mood of individuals \cite{canzian2015trajectories}.
In summary, the top five features for mood inference, regardless of whether it is two-class or three-class, were not the same across all countries. Certain features are unique to individual countries. At the same time, we can also observe a specific set of features (shown in the left quarter of both figures) that consistently appeared on the top list in all countries.

\section{Discussion}\label{sec:discussion}

In this section, we discuss the main findings of the paper, and highlight limitations and future work. 

\subsection{{{What do the Results Suggest?}}} 

In the country-specific setting, PLMs did not perform well across countries, with the highest performance for both two-class and three-class inferences coming from Mexico, with an AUROC of 0.62. However, performance increased significantly, with HMs showing the effect of personalization within countries. Comparable performance gains were observed for the multi-country setting as well. However, country-specific models (AUROC scores of 0.78-0.98 for two-class and 0.76-0.94 for three-class) would be preferred over multi-country models (0.83 and 0.79 for two and three classes, respectively). Then, in the country-agnostic setting, we observed that even HMs performed poorly compared to the country-specific setting. This means that if a model is trained in a different country, even if it is personalized to a person in another country, the model might not perform as well as a country-specific model that is personalized to a person in the same country. However, we also observed that models perform relatively better in culturally similar countries (i.e., within Europe). Within Asia, even though countries are in the same geographic region, cultural differences (i.e., India and China have different cultures and behaviors) could be one reason that did not allow models to perform better. Finally, building continent-specific models for Europe worked reasonably better than for Asia or a multi-country setting. Please note that the number of participants in several of these countries remained small, and so we cannot make any strong assertions.

\subsection{Comparison of Results to Previous Studies}

First, it should be noted that mood inference with smartphone sensing data is inherently a difficult task because of the task's subjectiveness. In this context, if we consider the results we obtained compared to some prior work, LiKamWa et al. \cite{likamwa2013moodscope} showed that they could achieve a 66\% accuracy with population-level models, around 75\% accuracy with hybrid models, and 94\% accuracy with user-level models. However, comparing the results in their paper to ours is difficult because we reported results with AUROC, which is a more holistic performance metric, especially in an imbalanced class scenario. However, purely in terms of numbers, the performance gain from PLM to HLM is greater in our case (from around 50\% to 80\%). This could be because of our dataset's more extensive set of features compared to their dataset, which only has phone usage-related features such as messages, calls, websites visited, and app usage. In addition, they modeled the inference as a regression task using multi-linear regression and provided model performance as a percentage using an error bound of 0.25 around the predicted value.

Another paper that used a similar dataset was by Servia-Rodriguez et al. \cite{servia2017mobile}. It is also worth noting that this dataset contains data from multiple countries, even though the analysis did not explicitly focus on that aspect. Furthermore, they only showed results for PLMs, obtaining an accuracy of around 70\% for weekends. Again, purely in terms of numbers, this is a good performance compared to what we obtained (AUROC scores of about 0.5). However, it is worth noting that they only reported results for weekends, for which inference performance was high, and we do not separate weekdays and weekends. In addition, the feature sets used for inference are again different. Another potential reason for the lack of performance in our PLMs could be participants' lack of movement during the pandemic when data were collected. This could result in sensors such as location (used in both the discussed papers) not being highly informative of different moods. Hence, this could lower the performance of our models. Interestingly, one common result across all three studies was that fewer negative labels were reported, which could make the development of fully personalized models more challenging due to the lack of data for negative classes from certain individuals. Hence, future studies could look into ways of capturing negative mood labels accurately and more often using different techniques. In addition, model personalization in situations where some users lack data for certain classes is a potential problem that could be explored further (a similar skewed labels-related scenario for depression detection has been discussed in a recent study \cite{xu2021leveraging}).

\subsection{{Diversity-Aware Research in Mobile Sensing}}

According to Gong et al. \cite{gong2019diversity}, diversity and diversity awareness are topics in machine learning that have gained importance in the recent past, and increasing generalization and decreasing biases in models for different populations are two fundamental goals discussed in this domain \cite{meegahapola2020smartphone}. According to them, diversity is achieved in machine learning with data diversification (maximizing the informativeness in training data such that the model fits data better), model diversification (increased diversity in model parameters leading to better learning), and inference diversification (model provides choices/information with more complementary information). Our study examined diversity awareness, primarily with data diversification. Since the whole data collection was done to emphasize the need for diversity awareness in machine learning-based mobile sensing systems, we defined diversity based on social practice theory \cite{harrison1998beyond, schelenz2021theory, giunchiglia2020diversity}. Accordingly, diversity is a complex and multi-layered construct that does not exist within individuals but surfaces when two or more individuals interact. Considering these conceptions, data and model diversity can be achieved by considering various types of diversity attributes ranging from country of residence, gender, and age, to personality, values, etc. \cite{harrison1998beyond, schelenz2021theory}. In this paper, we focused on ‘country of residence’ as an attribute for analysis because of the way mood is perceived and expressed, as well as phone usage and everyday behavior are different in countries around the world. In future work, other diversity attributes could be used to study mood (e.g., studying personality and mood with mobile sensing). Furthermore, other constructs collected in the study (e.g., social context, activity, food consumption) could be examined with mobile sensing, using country as a diversity attribute. 

\subsection{{Diversity-Awareness: Countries or Cultures?}}
In this study, we considered the geographical diversity of users when building smartphone sensing-based mood inference models. Hence, our primary construct of diversity is the `country of residence'. However, depending on the city, even though it is within the same country, the cultural composition of students could vary significantly. For example, our specific university in London, UK, is considered more diverse and has a high international student population compared to our specific university in India. These differences could also affect inference performance. In addition, our study also leaves the open question of whether the geographical region affects mobile sensing inference performance, or whether it is the culture of study participants that mediates their everyday life and phone usage behavior. Section~\ref{sec:inference} presented some initial results about these aspects. Future work could investigate these aspects further.

\subsection{{Ethical Considerations}}
Mood is a self-reported internal state and thus constitutes sensitive information. Ethical implications related to inference of affective states have been discussed in previous literature in affective computing \cite{cowie2015ethical, pai2022the, mohr2017personal}, ubicomp \cite{hilty2015ethical,nihan2013healthier}, and other disciplines \cite{bostrom2018ethics, mohammad2021ethics}. From the perspective of possible applications beyond supporting research on youth well-being, as we do here, it is fundamental that human-centered principles are followed and limit their use to cases that benefit individuals and avoid potential harm.

\subsection{{The Effect of the Pandemic and Weather on Mood Inference Models}}

In this paper, we showed how mood inferences could be done in the context of a mobile sensing application. In addition, we also showed how models lack generalization to unseen countries and the need for personalization. However, a limitation of this study is that the study was conducted during the pandemic. During the data collection time period in 2020, many countries have imposed different measures to curb the coronavirus. However, it is worth noting that, except for China, where strict lockdown measures were not present, universities have been in remote work/study mode in all the other countries. Hence, most students engaged in their studies from home. This could be the reason why there are many app usage, touch event, proximity, and wifi related features informative about mood according to Figure~\ref{fig:fi_two_class}, Figure~\ref{fig:fi_three_class}, and Table~\ref{tab:tstatistics}. It is also worth noting that the seasons in each country during the data collection period were different. On the positive side, none of the countries were in extreme winter or summer seasons. The September-November time period in European countries is the fall season, and none of those countries faced extreme cold weather conditions during that period. At this time, the season in Mongolia was comparable to European countries like Denmark or UK. All the other countries had comparatively higher temperatures. However, given that students in all the sites were affected by movement restriction measures and were stuck at home, we believe that weather conditions might not have affected the study as much compared to a time period when student behavior in outdoor environments would significantly change based on weather conditions. However, the results should be understood and interpreted with this limitation in mind. Future work could explore the effects of seasons and weather conditions on mobile sensing-based inferences.

\subsection{Domain Adaptation for Multi-Modal Mobile Sensing}
In this paper, we highlighted the issue of generalization and the possible distributional shifts in a mobile sensing dataset collected with the same protocol in different countries. Even though issues of generalization, biases, and domain shifts have been discussed extensively in other domains such as computer vision \cite{luo2019taking}, natural language processing \cite{elsahar2019annotate}, and speech \cite{sun2017unsupervised}, smartphone/mobile sensing studies have not focused on those aspects extensively thus far \cite{gong2021dapper}. Even though we provide evidence of the fundamental issue, we did not go into depth about finding a potential solution for that issue, as it is not within the scope of this paper (especially given page limits and extensive work that would be needed). Further, even though we showed that model personalization (hybrid setting) could minimize domain shift to an extent, other advanced techniques inspired by the work related to domain shift/adaptation in other domains could provide cues for solving such problems in mobile sensing. Recent studies also suggest that domain adaptation techniques for time series data are limited \cite{wilson2022domain}. For example, a longstanding problem in the human activity recognition (HAR) domain is the wearing diversity of wearables in different body positions. The wearing diversity hinders the performance of HAR models. A few recent studies suggested that unsupervised domain adaptation could be a solution for wearing diversity issues \cite{chang2020systematic, mathur2019unsupervised}.
Further, Wilson et al. \cite{wilson2022domain} explored domain adaptation for similar datasets captured from people from two age groups. However, the above studies focused on time series accelerometer data, which are more straightforward than the multi-modal datasets we are working with within this study. Hence, to the best of our knowledge, a research gap lies in solving domain adaptation for multi-modal sensing data coming from smartphones and wearables. In fact, in a recent study, Adler et al. \cite{adler2022machine} discussed the issue of generalization in multi-modal mobile sensing data and showed that lack of similarity across datasets collected in different time periods does not allow studying generalization of techniques to a greater depth. Therefore, with the dataset discussed in this paper, we believe solutions to domain adaptation and generalization could be explored further (not regarding generalization across time, but across geographically/culturally distinct areas), hence pushing the boundaries of multi-modal mobile sensing systems towards more real-world utility.

\subsection{Other Limitations and Future Work}

This work has several limitations and areas that could be improved in future work. First, the dataset used in this study is highly imbalanced, where there are fewer negative and very negative mood labels than neutral, positive, and very positive mood labels. However, this distribution is in a way similar to previous studies about valence \cite{servia2017mobile, likamwa2013moodscope}. Inherently, this also makes both inference tasks much harder. On the other hand, there is an imbalance in the dataset regarding data per country, where Italy and Mongolia had a significantly higher number of self-reports. In addition to the experimental results that we reported with imbalanced datasets, we conducted experiments with stratified down-sampled datasets for each country (each country having samples equal to the number of India, which had the lowest number of self-reports). While we reported some results for balanced cases in multi-country and continent-specific cases, more extensive analysis could be done to explore that aspect further. Hence, diversity-aware sampling strategies could be explored in future work to mitigate biases in mobile sensing-based inference models.
Further, we only considered valence in the circumplex mood model in this study. Other time diary questions were used to capture other behaviors and contexts, and we did not want to overburden users with multiple questions or lengthy questionnaires. However, we agree that collecting the arousal and understanding the geographical diversity of arousal inference could be studied in future work. In addition, the clinical validity of the valence in the circumplex mood models might be questionable. Future work could look into conducting studies with more clinically valid instruments for mood inference. In addition, in this paper, we did not use a 'wrapper' feature selection technique before training models because tree-based models, such as random forest, inherently use 'embedded' feature selection with Gini impurity to find a set of good features to build the trees with \cite{FeatureSelection2021}, especially when the feature space is small (i.e., around 100 in this dataset). However, if the feature space was larger, the dataset size was smaller, or if another non-tree-based model was used, using feature selection is highly preferred. Therefore, future work could also look into improving models based on feature selection and finding solutions to the issue of generalization using careful feature selection.

\section{Conclusion}\label{sec:conclusion}

In this exploratory study, we collected a mobile sensing dataset and around 329K self-reports from 678 participants in eight countries (China, Denmark, India, Italy, Mexico, Mongolia, Paraguay, UK) for over three weeks to assess the effect of geographical diversity on mood inference models. We evaluated country-specific, continent-specific, country-agnostic, and multi-country approaches trained on sensor data for two mood inference tasks with population-level (non-personalized) and hybrid (partially personalized) models. We showed that partially personalized country-specific models perform the best yielding AUROC scores in the range of 0.78-0.98 for two-class (negative vs. positive) and 0.76-0.94 for three-class (negative vs. neutral vs. positive) inference. Further, with the country-agnostic approach, we showed that models do not perform well compared to country-specific settings, even when models are partially personalized. We also uncovered generalization issues of sensing-based mood inference models to new countries. We hope that these findings will be of benefit to ubicomp researchers towards building future mobile sensing applications with an awareness of geographical diversity.

\begin{acks}

This work was funded by the European Union’s Horizon 2020 WeNet
project, under grant agreement 823783. We deeply thank all
the volunteers across the world for their participation in the study.

\end{acks}

\bibliographystyle{ACM-Reference-Format}
\bibliography{citations}

\appendix

\section{Appendix}

In this appendix, we describe how features were obtained for each sensing modality. This is an extension of the description we provided in Table~\ref{tab:agg-features}. First, it is worth noting that the analysis was done using a time window of 10 minutes. This would mean that for any sensing modality, we would filter out data for that particular time window. In addition to the sensor data within the time window, we also used the last data point before the start of the time window and the first data point after the end of the time window, in some cases when necessary.

\paragraph{\textbf{Location}} Location data were captured once every minute using either GPS signal or cell tower signals, depending on the most accurate signal available in a particular moment. We used the definitions for radius of gyration and distance covered from \cite{canzian2015trajectories}. This enables to capture the movement of the individual within the time window. Prior work has shown that movement could be indicative of different states related to mood and depression \cite{likamwa2013moodscope, servia2017mobile, canzian2015trajectories}. In addition, we also calculated the mean altitude using altitude values captured with location information.

\paragraph{\textbf{Bluetooth and Wifi}} There were two types of bluetooth devices logged with the mobile application. They are: low energy and normal. For each type, the mobile application logged a list of devices found with device IDs and signal strengths to each device. Then, we derived a set of features with a similar approach to \cite{santani2018drinksense}, including the number of device found, and statistical features regarding the signal strength to other devices in the vicinity. This sensing modality provides the user context as prior work has shown that these features could be indicative of whether the user is in a device-dense closed space or not \cite{santani2018drinksense, meegahapola2021one}. For Wifi, first, it was calculated whether the user is connected to a network or not. Usually someone connecting to a network indicates that they are in a familiar environment (i.e. home, workplace, university). In addition, similar to \cite{santani2018drinksense}, we also captured statistical features related to signal strength for all networks in the vicinity. This also provides data about the user context.

\paragraph{\textbf{Notifications}} The mobile app captured whenever users got a notification. In addition, in certain cases, unless the notification was clicked, the same notification would be displayed again (e.g. this could happen in WhatsApp). Hence, to capture these details, we calculated the number of notification posted by the system, and removed by the user, with and without the duplicates. This gives an indication of the phone usage behavior of users.

\paragraph{\textbf{Proximity}} Prior work has shown that proximity sensor reading could give an indication on where the phone is \cite{bae2017detecting}. Hence, basic statistical features were captured for the proximity variable. 

\paragraph{\textbf{Steps}} The step count was captured in the study using two techniques for reliability. First, the step count was derived using the total number of steps taken since the last time the phone was turned on. In addition, using a trigger in the system that sends an interrupt every time a new step is detected, the app also logged a separate step count called steps detected. We used both these features in the analysis. 

\paragraph{\textbf{Activity}} The mobile app provided the activity a person is doing, two times per minute. This activity was derived from a probability distribution of 8 activity types recognized by the Google Activity Recognition API. Therefore, we have a label of the activity the user is doing, roughly each 30 seconds. For example, if the first time window is from $T$ to $T+10$ mins, if the first activity label is at $T+1$ mins, the second activity report is at $T+2$ mins, we would assume that the user has been doing the first activity between $T+1$ min and $T+2$ min, for 1 minute. Similarly, we would calculate the approximate time of doing all the activities. However, it is worth noting that sometimes, the first activity label we get could be after 1-2 minutes after the start of the time window. This could happen because of inconsistencies in the data logged by the application. In such situations, we would also consider the last activity report before the start of the time window (let's say at $T-1$ min). We use the label of the last activity and include it in the calculation assuming that the user has been doing that activity from $T$ to $T+1$ min. Hence, using this technique, for each time window, we would have a distribution of activities the user has been doing in seconds. 

\paragraph{\textbf{Screen and Touch Events}} The mobile app logged whenever the screen was turned on or off, with timestamps. This allows us to calculate the time users spent with their screens turned on. For example, if the first time window is from $T$ to $T+10$ mins, if the screen was turned on at $T+3$ mins and turned off at $T+9$ mins, we could assume that the screen was turned off from $T$ to $T+3$ mins, and then it was on from $T+3$ to $T+9$. We consider 1 such turn on-off time period as an episode. A 10 minute time window could have multiple such episodes. Hence, using these values, we derived the number of episodes, statistical features for time spent in those episodes, and also the total time spent with the screen turned on. The distinction is that this allows us to distinctively identify a person who has the screen turned on for a longer duration vs. another person whose total screen on duration is the same as the earlier person, but turn on and off the screen more frequently (e.g. in a situation waiting for a email/message from someone, turning on the screen to see the time, etc.). We believe capturing these information could have value when studying attributes regarding mental well-being specially since screen on time has been associated to mental well-being in a lot of prior studies \cite{tang2021relationship}. In addition, the mobile app also logged all the touch events (this could be a tap or a keyboard press event). Using the values, we derived a feature for the total number of touch events in the time window.

\paragraph{\textbf{App Events}} Prior work that used app usage either considered the usage of individual apps \cite{meegahapola2021one} or app categories \cite{santani2018drinksense}. For this study, we felt that it's better to use app categories because of the heterogeneity of the dataset, where users from different countries would use different apps belonging to the same category, to do a similar task. We followed an approach similar to \cite{santani2018drinksense} and obtained the google play store app category (i.e. action, adventure, social, education, entertainment, etc.) for each app in the dataset, and used it to calculate app usage times. App usage time would be calculate from the time an app is on the screen to the time it closes or go to the background. There could also be instances where the phone screen is on and there are no apps on the screen. Such time periods were included in the category called "not\_found". In addition, whenever we could not find an app category for a particular app, it was also included under the "not\_found" category.

\end{document}